\documentclass[reprint,
 superscriptaddress,
 amsmath,
 amssymb,
 aps,
floatfix,
]{revtex4-2}

\usepackage{xpatch}

\makeatletter
\patchcmd{\@ssect@ltx}
{\addcontentsline{toc}{#1}{\protect\numberline{}#8}}
{}
{}
{}
\makeatother

\makeatletter
\auto@bib@empty   \makeatother

\usepackage{multibib}
\newcites{M}{Methods References}
\newcites{S}{Supplementary Information References}

\usepackage{array}[=2016-10-06]

\usepackage{graphicx}\usepackage{dcolumn}\usepackage{bm}\usepackage{hyperref, xcolor}\usepackage[all]{hypcap}
\usepackage{comment}
\usepackage{upgreek}
\usepackage{empheq}

\usepackage{booktabs}
\usepackage{colortbl}

\usepackage{etoolbox}
\preto\section{\par\addvspace{3ex}}
\preto\subsection{\par\addvspace{1.5ex}}

\newcommand{\LKB}{Laboratoire Kastler Brossel, Sorbonne Université, CNRS,
ENS-Université PSL, Collège de France, 4 place Jussieu, 75005 Paris, France}
\newcommand{\LPENS}{Laboratoire de Physique de l'Ecole Normale Supérieure, ENS-PSL, CNRS, Sorbonne Université, Université Paris Cité, Centre Automatique et Systèmes, Mines Paris, Université PSL, Inria, Paris, France}
\newcommand{\QUANTRO}{Quantronics group, Université Paris-Saclay, CEA,
CNRS, SPEC, 91191 Gif-sur-Yvette Cedex, France}

\newcommand{\AB}{Alice \& Bob, 53 Bd du Général Martial Valin, 75015 Paris, France}
\newcommand{\IUF}{Institut Universitaire de France (IUF), 1 rue Descartes, 75231 Paris, France}

\definecolor{niceblue}{RGB}{0, 80, 160}

\hypersetup{colorlinks, linkcolor=black, citecolor=niceblue, urlcolor=black}

\usepackage[T1]{fontenc}

\usepackage{xparse}
\newcommand{\op}[1]{\boldsymbol{#1}}
\newcommand{\sop}[1]{\boldsymbol{\mathcal{#1}}}
\newcommand{\ket}[1]{|#1\rangle}
\newcommand{\bra}[1]{\langle #1|}
\newcommand{\braket}[1]{\langle#1\rangle}
\newcommand{\nmth}{n_\mathrm{th}}

\newcommand{\sra}{{\scriptscriptstyle\rightarrow}}
\newcommand{\sla}{{\scriptscriptstyle\leftarrow}}
\newcommand{\opl}[1]{\op{{#1}}_\sra\!}
\newcommand{\opr}[1]{\op{{#1}}_\sla\!}
\newcommand{\rhop}{\op{\rho}}
\newcommand{\rhom}{\rhop_m}

\newcommand{\aop}{\op{a}}
\newcommand{\aopd}{\op{a}^\dag}
\newcommand{\tint}{\tau} 
\newcommand{\tcycle}{T}

\newcommand{\meanspec}[1]{\overline{\mathcal{S}}_{#1}}

\newcommand{\phiext}{\varphi_{\mathrm{ext}}}
\newcommand{\um}{~µm }

\newcommand{\mathum}{\upmu\mathrm{m}}

\newcommand{\mathnm}{\mathrm{nm}}

\newcommand{\Kr}{\sop{K}}
\newcommand{\Tr}{\mathrm{Tr}}
\newcommand{\II}{\mathbb{I}}

\newcommand{\Hms}{\op{H}_{\mathrm{ms}}}

\newcommand{\Veff}{V_\text{eff}}

\NewDocumentCommand{\reffig}{O{} m}{\hyperref[#2]{Fig.~\ref*{#2}#1}}
\NewDocumentCommand{\reffigfull}{O{} m}{\hyperref[#2]{Figure~\ref*{#2}#1}}
\RenewDocumentCommand{\refeq}{m}{\hyperref[#1]{Eq.~(\ref*{#1})}}
\NewDocumentCommand{\refeqfull}{m}{\hyperref[#1]{Equation~(\ref*{#1})}}
\NewDocumentCommand{\refsec}{m}{\hyperref[#1]{Sec.~\ref*{#1}}}
\NewDocumentCommand{\refapp}{m}{\hyperref[#1]{App.~\ref*{#1}}}
\NewDocumentCommand{\reftab}{m}{\hyperref[#1]{Tab.~\ref*{#1}}}
\NewDocumentCommand{\reftabfull}{m}{\hyperref[#1]{Table~\ref*{#1}}}

\newcommand{\nocontentsline}[3]{}
\newcommand{\tocless}[2]{\vspace{3ex}\par\bgroup\let\addcontentsline=\nocontentsline#1{#2}\egroup\par}

\makeatletter
\providecommand{\nocontentsline}{\relax}
\newenvironment{acknowledgments*}
{\bgroup
\let\addcontentsline=\nocontentsline
\begin{acknowledgments}}
    {\end{acknowledgments}\egroup
}
\makeatother

\usepackage[left]{lineno}

\setcitestyle{super,open=[,close=],sort&compress}

\usepackage{xr}
\usepackage{subfiles}

\begin{document}

\preprint{APS/123-QED}
\title{Probing the quantum motion of a macroscopic mechanical oscillator with a radio-frequency superconducting qubit}

\author{K. Gerashchenko}
\thanks{These three authors contributed equally}
\affiliation{\LKB}
\author{R. Rousseau}
\thanks{These three authors contributed equally}
\affiliation{\LKB}
\affiliation{\AB}
\author{L. Balembois}
\thanks{These three authors contributed equally}
\affiliation{\LKB}
\author{H. Patange}
\affiliation{\LKB}
\author{P. Manset}
\affiliation{\LKB}
\author{T. Briant}
\affiliation{\LKB}
\author{P.-F. Cohadon}
\affiliation{\LKB}
\author{A. Heidmann}
\affiliation{\LKB}
\author{W. C. Smith}
\affiliation{\LPENS}
\affiliation{Google Quantum AI, Santa Barbara, CA}
\author{A. Tilloy}
\affiliation{\LPENS}
\author{Z. Leghtas}
\affiliation{\LPENS}
\author{E. Flurin}
\affiliation{\QUANTRO}
\author{T. Jacqmin}
\thanks{These authors jointly supervised the work\\
samuel.deleglise@lkb.upmc.fr}
\affiliation{\LKB}
\affiliation{\IUF}
\author{S. Deléglise}
\thanks{These authors jointly supervised the work\\
samuel.deleglise@lkb.upmc.fr}
\affiliation{\LKB}

\date{\today}
 \maketitle

\textbf{
  Long-lived mechanical resonators like drums oscillating at MHz frequencies and operating in the quantum regime are a powerful platform for quantum technologies and tests of fundamental physics~\cite{Aspelmeyer2014}.
  Yet quantum control of such systems remains challenging, owing to their low energy scale and the difficulty of achieving efficient coupling to other well-controlled quantum devices. Here, we demonstrate repeated coherent interactions between a 4~MHz suspended silicon nitride membrane and a resonant superconducting heavy-fluxonium qubit~\cite{Manucharyan2009, Zhang2021, Najera2024}. The qubit is initialized at an effective temperature of 21~\textmu{}K and read out
  with 77\% single-shot fidelity. During the 6~ms lifetime of the membrane the two systems swap excitations more than 300 times. After each interaction, a
  state-selective qubit detection is performed, implementing a stroboscopic series of weak measurements that provide information about the mechanical state. The accumulated records reconstruct the position noise spectrum of the membrane, revealing both its thermal occupation $\bm{n_{\mathrm{th}} \approx 47}$ at 10~mK and the qubit-induced back-action. By preparing the qubit either in its ground or excited state before each interaction, we observe an imbalance between the emission and absorption spectra, proportional to $\bm{\nmth}$ and $\bm{\nmth + 1}$ respectively, a hallmark of the non-commutation of phonon creation and annihilation operators. Since the predicted Diósi–Penrose gravitational collapse time~\cite{Diosi1989, Penrose1996}  is comparable to the measured mechanical decoherence time, our architecture enters a regime where gravity-induced decoherence could be tested directly~\cite{Gely2021}.
}

\begin{figure*}[ht]
  \centering
  \includegraphics[width=\linewidth]{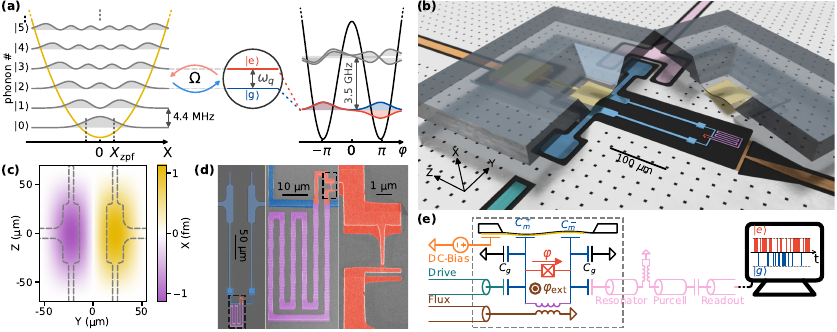}
  \caption{
    \textbf{Experimental principle and design.}
    \textbf{(a)} Energy-level diagram of the mechanical resonator (left) and fluxonium circuit (right).
    The membrane mode at frequency $\omega_m$ is modeled as a quantum harmonic oscillator.
    The heavy fluxonium’s potential (black line) at $\varphi_\mathrm{ext} = \pi$ is shown with its four lowest eigenstates.
    The qubit transition (blue/red) is tunable into resonance with the mechanical oscillator.
    \textbf{(b)} 3D schematic of the mechanical-fluxonium device. A superconducting loop composed of a small Josephson junction (red) and a superinductor (purple), is  threaded by an external flux $\varphi_{\rm ext}$ (brown). The small junction is shunted by a capacitor (blue). One capacitor electrode couples to a readout resonator (light pink) and the other to a control waveguide (cyan). The suspended SiN membrane, cut open for visibility, metallized with Al (yellow), and the two fluxonium electrodes form a vacuum‐gap capacitor. A charge gate (orange) provides DC bias and drives to the membrane.
    \textbf{(c)} Spatial profile of the $4.4\,$MHz mechanical mode of the membrane, plotted at an amplitude equal to its zero‐point fluctuations and overlaid on the fluxonium capacitor electrodes layout (dashed grey lines).
    \textbf{(d)} Colored optical and electronic micrograph of a twin fluxonium sample.
    \textbf{(e)} Lumped element circuit diagram of the sample.
    \label{fig:fig_design}
  }
\end{figure*}

Mechanical resonators, such as vibrating membranes or beams, are essential components of quantum technologies, serving as high-precision sensors~\cite{Rugar2004, Sansa2020, Gavartin2012, Catalini2020} and transducers between distinct quantum systems~\cite{Jiang2020, Andrews2014, Bagci2014}.
They also provide a platform for probing fundamental physics at the intersection of quantum mechanics and gravity~\cite{Gely2021}.
For instance, owing to their long lifetime and large zero-point fluctuations, state-of-the-art drums oscillating at MHz frequencies have been identified as good candidates to test the the gravitationally induced wavefunction collapse mechanism postulated by Diósi~\cite{Diosi1989} and Penrose~\cite{Penrose1996}.
Measuring and controlling the quantum state of these systems remains challenging because it requires coupling them to a resonant two-level system~\cite{Hofheinz2009}. In circuit Quantum Acoustodynamics (cQAD)~\cite{Arrangoiz2019, Bienfait2019, Chu2017}, this is typically achieved with superconducting qubits that operate in the GHz range.
Significant progress has been made toward bridging the frequency gap with macroscopic mechanical resonators. For instance, mechanical squeezing was demonstrated in a 22~MHz aluminum drum dispersively coupled to a 4~GHz Cooper-pair box~\cite{Viennot2018}.
However, coherent interaction and efficient qubit readout are hindered by the short coherence times of this qubit architecture.
More recently, techniques from cQAD have been extended to observe coherent exchange dynamics between a 700~MHz piezoelectric resonator and a fluxonium qubit~\cite{Lee2023}.
However, owing to the resonator’s extremely small mass (only a few picograms), the gravitational decoherence time predicted by the Diosi–Penrose model is six orders of magnitude longer than the environmental decoherence time.

In this paper, we report an experimental implementation of an idealized mechanical-cavity quantum electrodynamics system, where a mechanical resonator is repeatedly interrogated by a resonant two-level system.
The mechanical resonator is a suspended silicon nitride membrane oscillating at $\omega_m/2\pi \sim 4.4$~MHz.
It is coupled to a heavy-fluxonium circuit, which features simultaneously a low frequency qubit transition~\cite{Zhang2021, Najera2024}, resonant with the MHz mechanical mode, and GHz transitions to higher excited states that can be used to read out and prepare the qubit state deterministically (see \reffig[(a)]{fig:fig_design}).
We rely exclusively on the binary outcomes of qubit measurements to infer the mechanical state.
Owing to the long mechanical lifetime ($T_1^\mathrm{m} \sim 6$~ms), the qubit interacts with the membrane over 300 times during that interval,
extracting information bit-by-bit about its state.
This process realizes a periodic weak measurement of the mechanical mode, from which we reconstruct the noise spectrum $S_{xx}(\omega)$ of the dimensionless membrane position $\op{x}$ under the influence of the thermal environment of occupation $\nmth$.
By letting the membrane interact with the qubit either prepared in its ground, or excited state, we probe separately the emission and absorption noise spectra around $\pm \omega_m$.
This measurement reveals the fundamental quantum asymmetry between emission and absorption processes, proportional to $n_\mathrm{th}$ and $n_\mathrm{th} + 1$ respectively.
While similar signatures have already been observed through sideband asymmetries under dispersive coupling to microwave~\cite{Peterson2016} or optical~\cite{Safavi2012, Underwood2015, Xia2024} cavities, this is the first time, to our knowledge, the asymmetric mechanical noise spectrum is measured via coupling to a two-level system, as initially proposed theoretically~\cite{Clerk2010, Aguado2000, Gavish2000, Schoelkopf2003}.
Most importantly, given the mass of our system ($M\sim 5.3 $~ng), the estimated Diósi-Penrose gravitational collapse time~\footnote{This estimate relies on several debated assumptions; for a detailed discussion see~\refapp{sec:diosi_penrose}, Ref.~\cite{Gely2021} and references therein.} $\tau_G \sim 0.5$~ms is comparable to the mechanical decoherence time $\tau_\text{th} = 2T_{1}^m/n_{\textrm{th}}\sim 0.3$~ms, positioning our experiment in a regime where gravitational effects could become experimentally relevant~(see Methods). More generally, by bringing cQAD into the MHz regime, this platform opens the way to interferometric tests of objective-collapse models in the phenomenologically relevant parameter range (see Methods). \begin{figure*}[t]
    \centering
    \includegraphics[width=\linewidth]{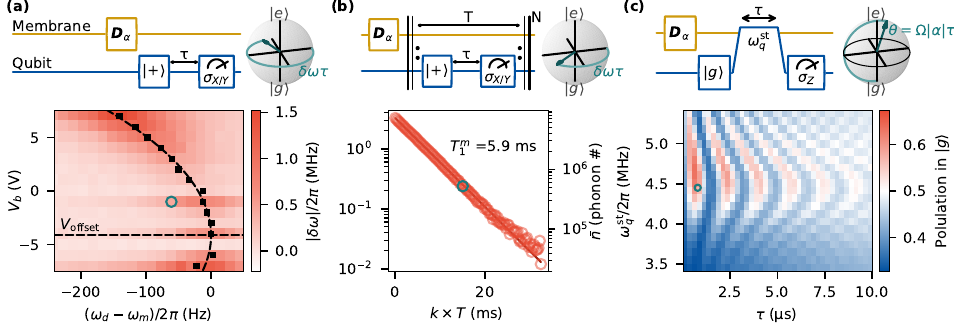}
    \caption{
        \textbf{Qubit-mechanical coupling.}
        \textbf{(a)} Ramsey spectroscopy under mechanical drive.
        \emph{Top:} Pulse sequence—each cycle begins with a mechanical displacement of the membrane, immediately followed by a fixed‐delay Ramsey sequence, repeated to sample the  qubit frequency shift \(\delta\omega = \chi\,\langle \hat a^\dagger \hat a\rangle\).
        \emph{Bottom:} Extracted qubit frequency shift versus bias voltage and drive detuning (color map). Dark squares mark the peak of the Lorentzian response at each \(V_b\), with a quadratic fit (dashed line). The drive amplitude is adjusted at each voltage to keep the peak response level approximately constant (except for $V_b = V_\text{offset}$ where the driving force vanishes).
        \textbf{(b)} Mechanical lifetime.
        \emph{Top:} Pulse sequence—after  resonant membrane displacement, a series of \(N=121\) Ramsey measurements are performed, sampling the instantaneous qubit frequency at discrete times \(k\times T\).
        \emph{Bottom:} Averaged qubit frequency shift \(\delta\omega\) plotted versus delay \(k\times T\) (for \(k=1,\dots,N\)), where data at each delay are ensemble‑averaged. An exponential fit yields a mechanical lifetime \(T_1^m = 5.9\) ms (\(Q = 1.62\times10^5\)).
        The right‑hand axis translates \(\delta\omega\) into the equivalent mechanical occupation in units of phonon number.
        \textbf{(c)} Coherent Rabi exchange.
        \emph{Top:} Pulse sequence—after initializing the membrane in a coherent state $\ket{\alpha}$, with $|\alpha|^2\simeq1.6\cdot10^5$, the qubit is Stark‐shifted to \(\omega_q^{\rm st}\) for an interaction time \(\tint\), and measured along \(\sigma_z\).
        \emph{Bottom:} Ground‐state population as a function of interaction time and Stark‐shifted qubit frequency. The spheres on the top right of each figure show the qubit Bloch vectors immediately after the interaction for the data points highlighted by blue circles.
    }
    \label{fig:interaction}
\end{figure*}

Our device (\reffig{fig:fig_design}) comprises two chips assembled via a flip‐chip technique. The fluxonium qubit is fabricated on intrinsic silicon (see Supplementary Information) and features two coplanar capacitor pads shunting a small Josephson junction (\(E_J/h = 4.82(6)\)\,GHz) in parallel with a large superinductance formed by 360 junctions (\(E_L/h = 0.128(7)\)\,GHz). The mechanical resonator is a clamped \(110\times140\,\mu\mathrm{m}^2\), 90-nm-thick SiN membrane under \(\gtrsim800\)\,MPa tensile stress, coated with a patterned Al pad (see Supplementary Information). The qubit couples predominantly to the membrane’s mode at \(\omega_m/2\pi = 4.4\)\, MHz—antisymmetric about the \(Y\)-axis (\reffig[(b)-(c)]{fig:fig_design})—via two vacuum-gap capacitors, yielding the total capacitance \(E_C/h = 0.418(9)\)\,GHz. A large bias electrode holds the membrane at potential \(V_b\), while capacitors \(C_g\) reference the fluxonium island to ground, enabling motion-to-charge transduction.

In the heavy‐fluxonium regime~\cite{Zhang2021, Najera2024} ($E_J \gg E_C > E_L$), the circuit is highly anharmonic, with a qubit transition frequency $\omega_{q}/2\pi = 2.35\,$MHz, while the transition to the next excited state occurs at 3.45~\text{GHz}. This permits treating $\{\ket{g},\ket{e}\}$ as an almost ideal two‐level system, with higher levels used for fast reset and dynamic tuning. A two‐photon preparation protocol yields fidelities $\eta_g = 99.5(1.7)\%$ and $\eta_e =99.2(5)\%$ (see Supplementary Information), while single‐shot dispersive readout achieves measurement fidelities $F_g = 84.8(1.8)\%$ and $F_e = 69.1(1.2)\%$ (see Supplementary Information). By driving the circuit near the higher excited state transitions at the flux sweet spot $\varphi_{\text{ext}}=\pi$, we dynamically tune $\omega_q$ via the AC Stark effect to bring it into resonance with the 4.4~MHz membrane. Coherence times are $T_1 = 22.4~$\textmu s and $T_2 = 17.1~$\textmu s, decreasing to $T_1 = 7.4~$\textmu s and $T_2 = 4.2~$\textmu s when tuned to $\omega_m$ via AC‐Stark shift (see Supplementary Information).

A membrane mode of frequency $\omega_m$ can be modeled as a harmonic oscillator of Hamiltonian $\op{H}_{m} = \hbar \omega_m (\op{a}^\dag \op{a} + 1/2)$, where $\op{a}$ and $\op{a}^\dag$ are the ladder operators  (\reffig[(a)]{fig:fig_design}).
Projecting the system dynamics in the $\{\ket{g}, \ket{e}\}$ manifold and applying the rotating wave approximation yields the Jaynes-Cummings Hamiltonian
\begin{equation}
    \op{H}_{\!\!J\!C}/\hbar =
    \omega_m \op{a}^\dag \op{a}
    + \frac{ \omega_q}{2} \op{\sigma}_z
    + i  \frac{\Omega}{2}  \left( \op{a}^\dag \op{\sigma} -  \op{a}\op{\sigma}^\dag \right),
\end{equation}
where $\op{\sigma}$ represents the lowering operator for the qubit and the coupling strength is given by (see  Supplementary Information)
\begin{equation}
    \Omega = \omega_q |\braket{g|\op{\varphi}|e}|\frac{\partial C_m^+}{\partial x} \beta \frac{ V_b - V_\text{offset}}{2e}.
\end{equation}
Here, \(V_b\) denotes the membrane bias voltage, $\beta \sim C_g/(C_g + C_m) \sim 1/2$ is a dilution factor resulting from the capacitive biasing scheme, and $V_\mathrm{offset} = Q_2/2 C_g$ is an offset voltage resulting from the static charge $Q_2$ confined on the floating fluxonium island.
With the maximum effective voltage of $|V_b - V_\mathrm{offset}| \sim 11.5$~V practically achievable in our setup, we measure $\Omega/2\pi \sim 1.50(8)$~kHz (see Supplementary Information).

 \begin{figure*}
    \centering
    \includegraphics[width=\linewidth]{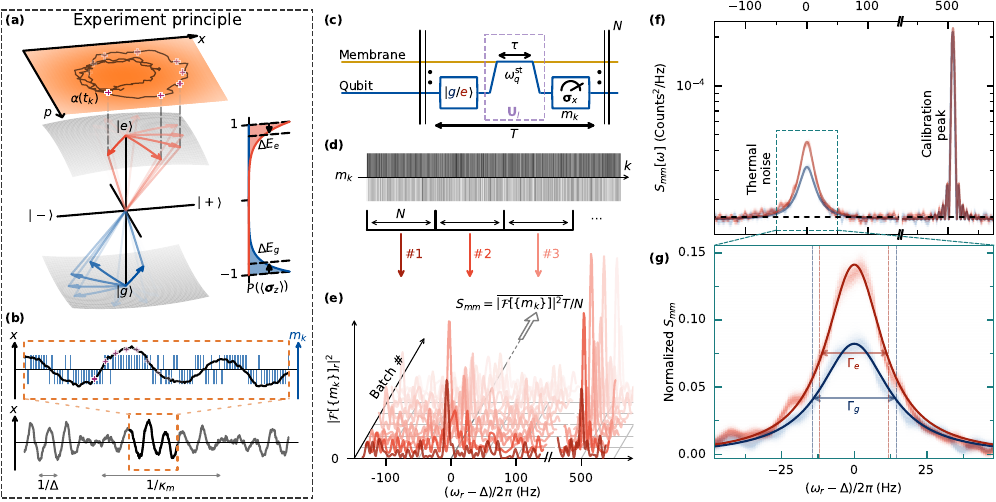}
    \caption{
\textbf{Spectrum analyzer experiment in the semi-classical limit.}
        \textbf{(a)} Phase-space distribution of the mechanical mode amplitude, with an overlaid classical trajectory (black curve), and the qubit Bloch sphere.
        After initialization in \(\ket{g}\) (blue) or \(\ket{e}\) (red), the membrane mode acts as a coherent drive on the qubit. The complex mechanical mode amplitude is mapped onto the  \((\op{\sigma_x}, \op{\sigma_y})\) plane of the Bloch sphere.
        The average energy transferred from membrane to qubit during the interaction   \(\Delta E_{g/e}\) is displayed on the $\langle\op{\sigma_z}\rangle$ histogram on the right‐hand inset.
        \textbf{(b)} Classical position \(x(t)\) of the mechanical resonator, shown in a frame rotating at the qubit–mode detuning \(\Delta\). The signal is an oscillation at frequency \(\Delta\), with amplitude and phase undergoing random fluctuations at rate \(\kappa_m=1/T_1^m\).
        Blue bars represent qubit \(\op{\sigma}_x\) measurement outcomes  after interaction with the mechanical mode.
        \textbf{(c)} Experimental pulse sequence.
        The qubit, prepared in $\ket{g}$ or $\ket{e}$, is Stark-shifted to a detuning $\Delta$ from the membrane frequency for a duration \(\tint\), followed by a $\op{\sigma}_x$ measurement.
        \textbf{(d)} Single‐shot outcomes \(m_k\in\{-1,1\}\) (for initial state \(\ket{e}\)) are recorded, grouped into batches of length \(N\), and each batch is discrete‐Fourier‐transformed. \textbf{(e)} Fourier transform of the first 10 batches represented as a waterfall plot.
        \textbf{(f)}  Position noise spectra obtained for the qubit prepared in \(\ket{g}\) (blue) and \(\ket{e}\) (red), by averaging the squared magnitudes of the Fourier transforms shown in (e). The two Lorentzian peaks on the left are centered at the membrane frequency and correspond to the membrane thermal noise. Additionally, a weak calibration tone produces a sinc‐shaped peak offset by 500~Hz towards higher frequencies.
        \textbf{(g)} Zoom on the Lorentzian peaks, with each spectrum normalized to the calibration‐tone amplitude and flat background subtracted. The broader, lower‐amplitude blue curve (\(\ket{g}\)-preparation) indicates qubit‐induced cooling, while the narrower, higher‐amplitude red curve (\(\ket{e}\)-preparation) indicates heating of the mechanical mode.
        Data in (a) and (b) are simulated for illustration purpose, whereas real data are displayed in (d), (e), (f), (g).
    }\label{fig:spectrum_analyzer}
\end{figure*}

In absence of AC-Stark effect, the qubit and the membrane modes are out-of-resonance: the large detuning $|\Delta| \gg \Omega$ suppresses direct energy exchange between the two systems.
Instead, the qubit undergoes a frequency shift proportional to the phonon population of the membrane~\cite{Gambetta2006, Koch2007} $\chi \op{a}^\dagger \op{a} \op{\sigma}_z$, where $\chi = \Omega^2/4\Delta$.
This effect is detected via a Ramsey spectroscopy experiment (see \reffig[(a)]{fig:interaction}).
The qubit is initialized in the $\ket{+}=(\ket{g} + \ket{e})/\sqrt{2}$ state and evolves for a duration $\tau$ with the membrane in a coherent state, during which it accumulates a phase proportional to the phonon number.
The experiment alternates between measuring the $\op{\sigma}_{x}$ and $\op{\sigma}_{y}$ components of the superposition over multiple realizations.
\reffig[(a)]{fig:interaction} shows the qubit frequency shift as a function of the membrane drive frequency and applied DC bias voltage $V_b$. The data reveal a quadratic decrease in the mechanical frequency with $V_b - V_{\rm offset}$, explained by the electrostatic spring-softening effect~\cite{Unterreithmeier2009}~(see supplementay information).

The relaxation time of the mechanical resonator $T_1^m$, is extracted by preparing the membrane in a coherent state and measuring the exponential decay of its phonon population.
This decay manifests as a time variation of the qubit frequency, tracked through repeated Ramsey measurements.
An exponential fit to the data, shown in \reffig[(b)]{fig:interaction}, yields $T_1^m = 5.9~\mathrm{ms}$.
The corresponding quality factor, $Q_m = \omega_m T_1^m = 1.62 \cdot 10^5$, is consistent with similar studies on silicon nitride membranes~\cite{Yu2012}.

We now focus on the resonant interaction regime.
In these experiments, the qubit is prepared and measured at its bare resonance frequency $\omega_q$, while the resonant interaction, is activated for a duration $\tint$ by tuning the qubit frequency $\omega_q^{\rm st} = \omega_m + \Delta$ close to the membrane one, governed by the evolution operator $\op{U}_{\!I} = e^{-i\op{H}_{\!\!J\!C}.\tau/\hbar}$.
When the membrane is initialized in a large enough coherent state $\alpha$, characteristic Rabi oscillations emerge at a rate $\Omega |\alpha|\gg 1/T_1$ (see \reffig[(c)]{fig:interaction}).
In this limit, $\op{U}_{\!I}$ describes a two-level system undergoing a rotation by an angle \(|\alpha| \theta_1 \) about the equatorial axis defined by $i \alpha^*$ under the influence of a classical drive.

When the membrane is in a thermal state, the resonant interaction induces a qubit precession of typical angle $\theta_1\sqrt{\nmth}$, where $\theta_1 = \Omega \tint \sim 47$~mrad is the angle induced by a single phonon.
Though small, this precession is detectable through sequential qubit measurements performed within a single mechanical lifetime~\cite{Cujia2019, Najera2024}.
Starting from the prepared qubit state (either \(\ket{g}\) or \(\ket{e}\)), we turn on the interaction for a duration $\tint=5~$\textmu s. We then project the qubit onto $\op{\sigma}_x$ by applying a $\sqrt{X}$ rotation,
followed by a $\op{\sigma}_z$ measurement.
This sequence is periodically repeated with the qubit reinitialized in the same state (\reffig[(c)]{fig:spectrum_analyzer}).

To build physical intuition, we first treat the degrees of freedom of the membrane—its position $\op{x}$ and momentum $\op{p}$—classically.
In the qubit drive frame, rotating at $\omega_q^{\rm st}$, the complex mechanical amplitude $\alpha(t) = \tilde \alpha(t) e^{i \Delta t}$ appears as a signal oscillating at frequency $\Delta$, modulated by a slowly varying envelope $\tilde \alpha(t)$ that diffuses in phase space at a characteristic rate $\kappa_m$, as shown in~\reffig[(a)]{fig:spectrum_analyzer}.
This amplitude is sampled at discrete times $t_k = k \times \tcycle$ by the sequential qubit measurements of \( \op{\sigma}_{x, k}\), where $\tcycle = 15.5~$\textmu s. In the small angle limit $|\alpha|\theta_1\ll 1$, the rotation $\op{U}_{\!I}$ maps
the phase space coordinates \((\mathrm{Re}(\alpha(t_k)), \mathrm{Im}(\alpha(t_k))\) of the mechanical oscillator onto the qubit's transverse components \((\op{\sigma}_{x, k}, \op{\sigma}_{y,k})\) (\reffig[(b)]{fig:spectrum_analyzer}). Therefore,
\begin{equation}
  \label{eq:sigma_x_mean}
  \braket{\op{\sigma}_{x,k}} = \theta_1 x(t_k),
\end{equation}
where \(x(t_k) =  \mathrm{Re}(\alpha(t_k))\) is the dimensionless membrane position.
By operating the qubit in a frame detuned by $\Delta > \kappa_m$, the free evolution between interaction pulses continually rotates the measurement axis. Consequently, successive $\op{\sigma}_{x, k}$ measurements sweep through all mechanical quadrature angles—probing both $\mathrm{Re}(\tilde \alpha(t_k))$ and $\mathrm{Im}(\tilde \alpha(t_k))$—in direct analogy with a heterodyne detection.

Our goal is to characterize the steady‐state noise spectra \(S_{mm}^{g/e}(\omega)\) of the qubit readout record \(m_k\). In practice see Methods), we segment the time series into equal‐length batches, compute the discrete Fourier transform of each, and average the squared magnitudes (see \reffig[(d)-(e)]{fig:spectrum_analyzer}).
In a semiclassical approximation that neglects qubit backaction on the membrane, the trajectory \(x(t)\) acts as an independent external drive, and the two‐point correlator factorizes:
\begin{equation}
  \label{eq:autocorrelator-factorization}
  \mathcal{C}_k \equiv \overline{\langle\sigma_{x,k}\,\sigma_{x,0}\rangle}
  =
  \overline{\langle\sigma_{x,k}\rangle}\,\overline{\langle\sigma_{x,0}\rangle},
\end{equation}
for $k\neq0$, with $\mathcal{C}_0 = 1$.
Recalling \refeq{eq:sigma_x_mean}, and Wiener–Khinchin theorem, linking the measured spectrum to the Fourier-transform of $\mathcal{C}_k$, we have
\begin{equation}
  \label{eq:spectra_semi_classical}
  S_{mm}^{g/e}(\omega) = \frac{T}{N}\mathcal{F}[\mathcal{C}_k]=\frac{\theta_1^2}{4} S_{xx}(\omega) + \tcycle.
\end{equation}
where \(S_{xx}\) is the position noise spectrum and \(\tcycle\) is the constant background arising from the sampling noise due to the binary nature of the qubit measurements.
In our experimental data (see \reffig[(f)]{fig:spectrum_analyzer}), the spectra $S_{mm}^{g/e}$ display a Lorentzian peak centered at $\omega_m$ that reflects the thermal fluctuations of the membrane, with the sampling noise background $\tcycle$ indicated by a dashed line.
Although \refeq{eq:spectra_semi_classical} predicts identical spectra regardless of the initial qubit state, we observe significant differences in both the area and the linewidth of the thermal peaks.

These discrepancies arise because our simple model neglects the energy exchange between the membrane and the qubit during their interaction. In particular, when the qubit is prepared in $\ket{g}$ it absorbs energy from the membrane, whereas
when prepared in $\ket{e}$, it emits energy to the membrane (see \reffig[(a)]{fig:spectrum_analyzer}).
On average, the qubit energy is lowered or raised on the Bloch sphere by $\Delta E_{g,e}/\hbar \omega_q^\text{st} =  \theta_1^2 \nmth/4$ due to the interaction.
Because of energy conservation, the membrane must experience a qubit-induced relaxation or amplification at a rate $\kappa = \theta_1^2/4 \tcycle$.
The mechanical linewidth is thus modified (broadened or narrowed) according to $\kappa'_{g/e} = \kappa_m \pm \kappa$.
Furthermore the qubit being reinitialized in a pure state before each interaction, it provides a 0-occupation bath for the membrane, leading to an occupation $n'_{\mathrm{th}, i}= \kappa_m\nmth /\kappa'_i$.
This qubit-induced cooling (or heating) mechanism is analogous to the dynamical back-action effects observed in cavity optomechanics~\cite{Arcizet2006, Schliesser2006}.

Lorentzian fits to the thermal peaks yield linewidths of $\kappa'_g = 2\pi\times(28.5\pm 1.4)$~Hz for $\ket{g}$ preparations and $\kappa'_e = 2\pi\times(23.7\pm 0.7)$~Hz for $\ket{e}$ preparations.
The inferred linewidth $\kappa_m = (\kappa'_e + \kappa'_g)/2 = 2\pi \times (26.1\pm0.8)~{\rm Hz}$ is in agreement with the previously measured $1/T_1^m = 2\pi\times(27.16\pm 2.6)$~Hz, confirming that the membrane exhibits no additional dephasing beyond its Fourier-limited linewidth.
Moreover, the dynamical back-action rate $\kappa = 2\pi\times(2.4\pm0.8)$~Hz agrees with a full experimental model that includes finite qubit preparation and readout fidelities.
To account for these imperfections, the spectrum is normalized using a calibration peak generated by a weak continuous tone applied on the qubit charge drive port (see \reffig[(f)]{fig:spectrum_analyzer}).
The normalized, background-subtracted spectra are displayed in \reffig[(g)]{fig:spectrum_analyzer}.
The experimentally obtained area ratio $A_e/A_g = 1.42 \pm 0.11$ is in a reasonable agreement with the expected value $n'_{\mathrm{th}, e}/n'_{\mathrm{th}, g} =  \kappa'_g/\kappa'_e = 1.22 \pm 0.07$.
Here the dynamical backaction is moderate due to the relatively low cooperativity \(\kappa/\kappa_m \sim 1/10\).
Nevertheless, we expect that ground‐state cooling will be within reach in future experiments—either by further increasing the mechanical lifetime~\cite{Tsaturyan2017, Ivanov2020} or by boosting the qubit–membrane coupling strength via a larger DC bias.

 \begin{figure}
    \centering
    \includegraphics[width=\linewidth]{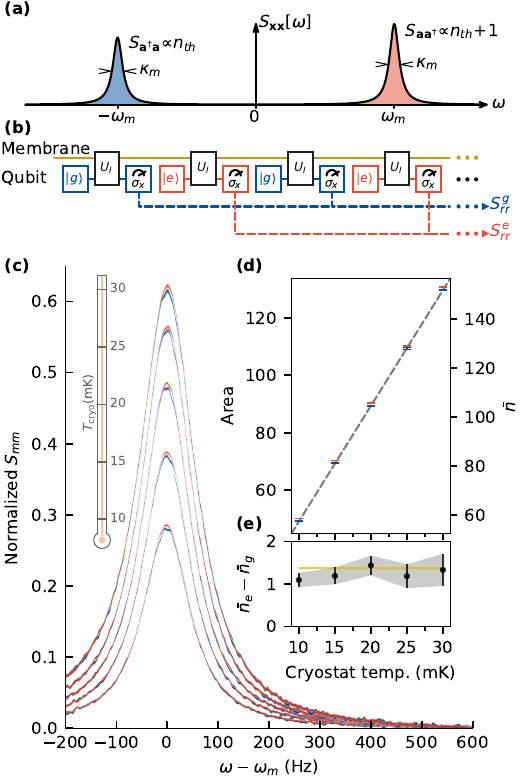}
    \caption{\textbf{Quantum position-spectrum asymmetry.}
        \textbf{(a)} Position‐noise spectrum of a harmonic oscillator at thermal equilibrium, showing absorption (blue peak at \(-\omega_m\), area \(\propto \bar n_{\rm th}\)) and emission (red peak at \(+\omega_m\), area \(\propto \bar n_{\rm th}+1\)).
        \textbf{(b)} Experimental pulse sequence: to eliminate the average dynamical backaction, we alternately initialize the qubit in the ground state \(\ket{g}\) and the excited state \(\ket{e}\). From the resulting measurement records, we then reconstruct the emission spectrum \(S_{xx}^g\) (when starting in \(\ket{g}\)) and the absorption spectrum \(S_{xx}^e\) (when starting in \(\ket{e}\)).
        \textbf{(c)} Reconstructed spectra for \(\ket{g}\) (blue) and \(\ket{e}\) (red) preparations at cryostat temperatures of 10, 15, 20, 25, and 30~mK, as illustrated by the thermometer.
        \textbf{(d)} Lorentzian‑peak areas at \(\omega_m\) versus cryostat temperature: blue and red symbols show the peak areas in \(S_{xx}^g\) and \(S_{xx}^e\) respectively.
        A linear fit to the mean-area of the 5 temperature series is shown (dashed line), and the right‑hand axis converts the fitted peak area into mean phonon number using this calibration.
        \textbf{(e)}  Measured phonon‐number difference between \(\ket{e}\) and \(\ket{g}\) spectra versus cryostat temperature. The horizontal dashed line at 1.37 quanta is the theoretical prediction, which differs from 1 due to qubit preparation infidelity and finite lifetime. All points include error bars representing the standard deviation from bootstrap fits to random FFT‐batch sub‐ensembles (Sec. S4.6  of Supplementary Information).
    }
    \label{fig:spec_asym}
\end{figure}

In the full quantum treatment, each interaction cycle constitutes a weak measurement of the membrane state \(\op{\rho}_m\), described by
\[
    \op{M}_\pm^i = \bra{\pm}\,\op{U}_I\,\ket{i},\quad i\in\{g,e\},
\]
where \(\op{U}_I\) entangles qubit and membrane and “\(\pm\)” labels the projective qubit outcome.  This measurement‐induced backaction steers the membrane into one of two conditional states, imprinting information that survives for its mechanical lifetime \(T_{1}^m\). As a result, for qubit prepared in \(\ket{g}\) (resp.\ \(\ket{e}\)), the two‐point correlator to second order in \(\theta_1\) becomes:
\[
    C_k^g = \theta_1^2\,\text{Re}\bigl\langle \aopd(t_k)\,\aop(0)\bigr\rangle,
    \qquad
    C_k^e = \theta_1^2\,\text{Re}\bigl\langle \aop(t_k)\,\aopd(0)\bigr\rangle,
\]
where $\aop(t)$ and $\aopd(t)$ are the annihilation and creation operators in the Heisenberg picture. Fourier transforming these correlators yields the emission (\(g\)) and absorption (\(e\)) spectra of the mechanical mode (see Methods).

In \reffig{fig:spectrum_analyzer}, the  difference between $S_{xx}^{g}(\omega_m)$ and $S_{xx}^{e}(\omega_m)$ is largely due to classical backaction--cooling for $\ket{g}$ and amplification for $\ket{e}$.
In other words, the equilibrium occupation $\langle\aopd \aop  \rangle$ is different in the two experimental sequences.
To mitigate this effect, and isolate the small quantum asymmetry, we implemented an  experimental sequence where the qubit is alternatively prepared in $\ket{g}$ and $\ket{e}$ (see \reffig[(b)]{fig:spec_asym}). In this manner, the average energy exchanged with the qubit cancels out in a coarse-grained approximation.
The spectra $S_{mm}^{g/e}$ are then computed from the corresponding subsequences $\{m^{g/e}_k\}$ (see \reffig[(b)]{fig:spec_asym}).

The experimental spectra $S_{mm}^{g/e}$, represented in blue and red in \reffig[(c)]{fig:spec_asym}, exhibit a small but statistically significant difference in their areas.
To confirm that this offset has a quantum origin, we vary the cryostat temperature, $T_\mathrm{cryo}$, within the interval $[10\,\mathrm{mK}, 30\,\mathrm{mK}]$.
The spectra are normalized using a calibration peak and corrected by subtracting the sampling noise background.
The thermalization of the mechanical mode with the cryostat is demonstrated by the linear dependence of the mean peak area.
From this linear dependence, we infer the mechanical mode temperature as $T_\mathrm{m} = T_\mathrm{cryo} + 2.2$~mK, allowing us to recalibrate the mode area in units of quanta (right axis of \reffig[(d)]{fig:spec_asym}).
The constant offset of $2.2$~mK observed can be attributed either to a miscalibration of the generic Bluefors temperature sensor, additional acoustic noise transduced into the mechanical mode, or a slight temperature gradient between the sensor location near the mixing chamber and the sample holder.
Most notably, the very small yet statistically
significant offset between the peak areas in $S_{mm}^{g}$  and $S_{mm}^{e}$, is constant across the entire temperature range, with an average of 1.25(25) phonons.
This experimental offset—slightly above the ideal quantum asymmetry of exactly 1 phonon—agrees well with the 1.37-phonon prediction of our full theoretical model, which incorporates finite state-preparation fidelity and qubit-lifetime (see Supplementary Information).

 To conclude, we have demonstrated, for the first time, direct resonant coupling between a superconducting qubit and a macroscopic mechanical membrane oscillating at 4 MHz—over two orders of magnitude below previous realizations~\cite{Lee2023}. By employing a heavy-fluxonium architecture, we bridge the gap between the GHz-scale of superconducting circuits and the low-frequency mechanical domain, while preserving coherence. Exploiting the membrane’s long lifetime and periodic weak measurements, we reconstruct its noise spectrum and reveal a clear asymmetry between emission and absorption—a direct signature of the non-commutativity of phonon creation and annihilation operators.

Importantly, operating a \(5.3\)\,ng membrane at 4~MHz brings us into the regime where a macroscopic spatial superposition might decohere via the Diósi–Penrose mechanism on timescales comparable to the intrinsic coherence of the membrane~\cite{Gely2021}.
Looking ahead, this platform naturally lends itself to quantum and adaptive feedback protocols—using membrane drives or conditional qubit preparations informed by real-time readout—to achieve active stabilization, state steering, and generation of non-classical mechanical states~\cite{Sayrin2011}.
Several optimizations could improve the device performance. Phononic‐crystal engineering should boost the electromechanical cooperativity by three orders of magnitude. Increasing the DC bias above 11.5~V—by mitigating leakage currents—, reducing the membrane–qubit separation to sub-µm scales, and applying standard surface passivation to extend the qubit coherence time \(T_2\) will together maximize the coupling–coherence product \(\Omega T_2\). These measures should close the current order-of-magnitude gap and enable operation in the strong-coupling regime, thereby enabling experimental tests of wavefunction collapse models.

\bibliographystyle{naturemag}

{\let\oldaddcontentsline\addcontentsline
    \renewcommand{\addcontentsline}[3]{}\bibliography{bib-main}
    \let\addcontentsline\oldaddcontentsline
}

\renewcommand{\thesubsection}{\Alph{subsection}}
\renewcommand{\thesubsubsection}{\thesubsection.\arabic{subsubsection}}
\makeatletter
\renewcommand{\p@subsection}{}
\renewcommand{\p@subsubsection}{}
\makeatother

\tocless{\section*}{Methods}

\tocless{\subsection}{Device Parameters\label{sec:parameters_table}}

\reftabfull{tab:parameters} summarizes the key device parameters; the fourth column indicates the characterization method—often citing the specific section where each measurement is described.
During the interval between the measurements for \reffig{fig:spectrum_analyzer} and \reffig{fig:spec_asym}, an unexpected temperature rise in the dilution refrigerator reduced the mechanical lifetime from $T_1^m = $~5.8~ms to $T_1^m = 1.47$~ms. Concurrently, the interrogation time was shortened to 4~µs, yielding a single phonon rotation angle of \(\theta_1 = 38(2)\)~mrad.

\begin{table*}[h]
    \centering
    \normalsize
    \begin{tabular}{l l l l}
        \hline
        Parameter        & Description                                     & Value              & Origin                                                     \\
        \hline
        \multicolumn{4}{c}{\textbf{Membrane}}                                                                                                                \\
        \hline
        $\omega_m/2\pi$  & Membrane frequency                              & 4.4 MHz            & Membrane spectroscopy \reffig[(a)]{fig:interaction}        \\
        $M$              & Physical mass                                   & 5.3 ng             & \refsec{sec:effective_mass} of S.I.                        \\
        $m$              & Effective mass                                  & 2.3 ng             & \refsec{sec:effective_mass} of S.I.                        \\
        $X_\text{zpf}$   & Zero-point fluctuations                         & 0.9 fm             & \refsec{sec:circuit_quantization} of S.I.                  \\
        $T_1^m$          & Mechanical lifetime                             & 5.9 ms             & Ringdown experiment \reffig[(b)]{fig:interaction}          \\
        $Q$              & Mechanical quality factor                       & $1.62\times10^5$   & Ringdown experiment \reffig[(b)]{fig:interaction}          \\
        $\tau_G$         & Expected gravitational collapse time            & 0.5~ms             & \refsec{sec:diosi_penrose}                                 \\
        $\tau_\text{th}$ & Thermal decoherence time                        & 0.3~ms             & \refsec{sec:diosi_penrose}                                 \\
        \hline
        \multicolumn{4}{c}{\textbf{Qubit}}                                                                                                                   \\
        \hline
        $E_J/h$          & Josephson energy                                & 4.82(6) GHz        & Spectroscopy fit (\refsec{sec:qubit-spectroscopy} of S.I.) \\
        $E_L/h$          & Superinductance energy                          & 0.128(7) GHz       & Spectroscopy fit (\refsec{sec:qubit-spectroscopy} of S.I.) \\
        $E_C/h$          & Charging energy                                 & 0.418(9) GHz       & Spectroscopy fit (\refsec{sec:qubit-spectroscopy} of S.I.) \\
        $\omega_q/2\pi$  & Qubit $\ket{g}\leftrightarrow\ket{e}$ frequency & 2.35 MHz           & Figure~S7 (Sec.~S3.4 of S.I.)                              \\ $T_1$            & Qubit $T_1$ (bare)                              & 22.4(8) \textmu s  & Fig.~S8 (\refsec{sec:qubit-decoherence} of S.I.)           \\
        $T_2$            & Qubit $T_2$ (bare)                              & 17.1(1) \textmu s  & Fig.~S8 (\refsec{sec:qubit-decoherence} of S.I.)           \\
        $T_1^{\rm st}$   & Qubit $T_1$ (Stark-shifted)                     & 7.4(2) \textmu s   & Fig.~S8 (\refsec{sec:qubit-decoherence} of S.I.)           \\
        $T_2^{\rm st}$   & Qubit $T_2$ (Stark-shifted)                     & 4.2(0) \textmu s   & Fig.~S8 (\refsec{sec:qubit-decoherence} of S.I.)           \\
        $\eta_g$         & Preparation fidelity $\ket{g}$                  & 99.5(1.7)\%        & \refsec{sec:qubit-preparation} of S.I.                     \\
        $\eta_e$         & Preparation fidelity $\ket{e}$                  & 99.2(5)\%          & \refsec{sec:qubit-preparation} of S.I.                     \\
        $F_g$            & Readout fidelity $\ket{g}$                      & 84.8(1.8)\%        & \refsec{sec:qubit-readout} of S.I.                         \\
        $F_e$            & Readout fidelity $\ket{e}$                      & 69.1(1.2)\%        & \refsec{sec:qubit-readout} of S.I.                         \\
        \hline
        \multicolumn{4}{c}{\textbf{Coupling}}                                                                                                                \\
        \hline
        $d$              & Membrane–qubit gap distance                     & $2.5(3)~$\textmu m & \refsec{sec:qubit-membrane_distance} of S.I.               \\
        $C_m(d)$         & Membrane–qubit capacitance                      & 13.9 fF            & Ansys simulation                                           \\
        $\Omega/2\pi$    & Qubit–membrane vacuum rabi rate                 & 1.50(8) kHz        & \refsec{sec:qubit_membrane_coupling} of S.I.               \\
        $\theta_1$       & Single-phonon rotation angle                    & 47(3) mrad         & From $\Omega$                                              \\
        $C_q$            & Qubit shunt capacitance                         & 5.7 fF             & Ansys simulation                                           \\
        $C_g$            & Qubit–ground capacitance                        & 44.3 fF            & Ansys simulation                                           \\
        \hline
    \end{tabular}
    \caption{Key device parameters, grouped by Membrane, Qubit, and Coupling. Uncertainties are reported for experimentally determined parameters when they exceed the resolution of the last quoted digit.}
    \label{tab:parameters}
\end{table*}
 \tocless{\subsection}{Fundamental tests on massive quantum superpositions\label{sec:fundamental_test}}

According to general relativity, mass–energy curves spacetime, so the rate at which time flows is not universal but depends on the distribution of nearby massive objects.
On the other hand, in quantum mechanics, the superposition principle allows a system to occupy several distinct configurations at once.
This leads to a conceptual tension: how does time flow for a massive object in a spatial superposition? In Schrödinger’s equation, time appears only as an external parameter, independent of the quantum state.

Concomitantly, while the superposition principle has been verified extensively at the microscopic scale, it is in stark contrast with our everyday experience where an object can only be in one specific state at a time.
The emergence of a classical macroscopic world is not derived from the Schrödinger equation but introduced by the measurement postulate: during a measurement the unitary evolution ceases and the state is projected onto a single outcome.
Neither the moment at which unitary evolution gives way to projection nor the reason why measuring devices are singled-out is specified by the theory; this is the quantum measurement problem.

Spontaneous-collapse models resolve this problem by replacing the  measurement postulate with a modification of Schrodinger's equation. Added nonlinear noise terms steer spatial superpositions continuously toward localized states, with a strength that scales with the system’s mass or complexity. Consequently, the quantum dynamics of microscopic systems is practically unchanged, while macroscopic superpositions are suppressed on experimentally negligible timescales. Such modifications would explain the quantum–to-classical crossover within one unified dynamical law.
Importantly, the collapse dynamics, controlled by a few model parameters, differs from that of standard quantum mechanics, such that it can be verified experimentally. Among these models, the Diósi–Penrose proposal~\citeM{Penrose1996,Diosi1989} plays a central role by explicitly linking collapse to gravity, with self-gravity setting the timescale for the decay of spatial superpositions. In other words, massive superpositions are dynamically localized so that spacetime remains effectively classical.

Efforts to bound collapse-model parameters follow two complementary strategies. \emph{Interferometric} tests are the most direct: they monitor the loss or persistence of fringe visibility as ever more massive objects are prepared in spatial superpositions. \emph{Non-interferometric} tests, by contrast, search for indirect signatures of the same localization noise—momentum diffusion, excess heating, or anomalous radiation—at the cost of additional assumptions about environmental couplings, noise spectra, and electromagnetic response.

So far, the strongest interferometric bounds have come from a bottom-up program in molecular and cold-atom matter-wave interferometry, reaching object masses of order $10^4$ atomic mass units (amu).
In parallel, stringent non-interferometric limits have been obtained in solid-state platforms by monitoring excess Brownian noise, from cryogenic cantilevers to the mirrors of gravitational-wave detectors. For such optomechanical systems, however, the main obstacle to a direct interferometric test has been the preparation of genuine quantum superpositions. Recent progress in circuit quantum acoustodynamics (cQAD) changes this outlook, enabling a top-down route in which macroscopic solid-state resonators can be prepared in nonclassical motional states, with masses beyond $10^{15}$ amu.
In particular, a 16~µg resonator has recently been prepared in a Schrödinger-cat state~\citeM{Bild2023}.
Up to now, however, most cQAD platforms have operated in the hundreds-of-MHz to GHz range, which directly constrains the accessible spatial separation. For a Schrödinger-cat state
\begin{equation}
    \label{eq:cat}
    \ket{\Psi} = \frac{1}{\mathcal{N}} (\ket{\alpha} + \ket{-\alpha}), \qquad (\mathcal{N}\approx \sqrt{2}),
\end{equation}
the distance between the two coherent components of amplitude $\alpha$ is
\[
    d = 2|\alpha|\,X_{\rm zpf},
    \qquad\text{with}\quad
    X_{\rm zpf}=\sqrt{\frac{\hbar}{2m\omega_m}},
\]
so at fixed $m$ and $\alpha$ one has $d\propto\omega_m^{-1/2}$.
With the parameters of Ref.~\citeM{Bild2023} and $|\alpha|^2 = 2.6$, the corresponding separation is $d \approx 1$~am.
In our MHz regime $X_{\rm zpf}\approx 1~\mathrm{fm}$, allowing separations of a few femtometres for
$|\alpha|^2\sim 1 -10$. These femtometre-scale separations match nuclear dimensions—the scale first invoked in collapse proposals—and  bring such models within direct interferometric reach.
We review below the main collapse models, namely the Diosi-Penrose and CSL frameworks and assess the bounds on the model parameters achievable with our platform.

\tocless{\subsubsection}{cQAD tests of objective collapse models}

Although collapse models are often formulated as stochastic, nonlinear modifications of the Schrödinger equation for state vectors, it is convenient to describe the averaged dynamics at the density-matrix level. In this picture, the collapse appears as an additional decoherence term in the Lindblad master equation:
\begin{equation}
    \dot{\op{\rho}} \;=\; \sop{L}_{\rm QM}[\op{\rho}] \;+\; \sop{L}_{\rm collapse}[\op{\rho}] .
\end{equation}
Requiring linearity, complete positivity, translation invariance, and no superluminal signaling tightly constrains its form. A general Markovian expression is
\begin{equation}
    \label{eq:collapse_kernel}
    \sop{L}_{\rm collapse}[\op{\rho}]
    = \,\!\int d^{3}\vec r\,d^{3}\vec r'\;
    \mathcal{D}(\vec r-\vec r')\,
    \big[\op M(\vec r),\,[\op M(\vec r'),\,\op{\rho}]\big],
\end{equation}
\noindent
where $\op M(\vec r)$ denotes the mass-density operator and $\mathcal D$ encodes the strength and spatial correlations of the localization field. The double commutator suppresses off-diagonal matrix elements of
$\op{\rho}$ between states with different mass-density profiles. The kernel $\mathcal{D}(\vec r)$ specifies the overall strength and spatial resolution of this collapse (only mass-density differences on scales larger than this width are efficiently resolved). Different collapse proposals correspond to different choices of $\mathcal{D}$. We will specify these choices below for the Diósi–Penrose and CSL models.

In an interferometric cQAD test, a mechanical Schrödinger-cat state \(\ket{\Psi}\) with calibrated  separation \(d\) [\refeq{eq:cat}] is prepared and its coherence is tracked in time. On this platform, the mechanical phonon-number parity, provides a direct readout of fringe visibility, which—under Markovian conditions—decays exponentially with rate
\begin{equation}
    \Gamma_{\rm meas}(d)=\Gamma_{\rm QM}(d)+\Gamma_{\rm collapse}(d;\theta).
    \label{eq:budget_interf}
\end{equation}
Here \(\Gamma_{\rm QM}(d)\) denotes the standard decoherence rate predicted by quantum mechanics from environmental couplings (e.g.\ thermal noise and damping, in the absence of any collapse mechanism), while \(\Gamma_{\rm collapse}(d;\theta)\) is the model-dependent excess;
By comparing the measured decoherence rate $\Gamma_{\rm meas}(d)$ to the independently calibrated value $\Gamma_{\rm QM}(d)$, one can bound $\Gamma_{\rm collapse}(d;\theta)$ and thus exclude some regions of the parameter's model $\theta$.
A key advantage of cQAD is that the cat separation \(d\) is precisely set by state preparation, allowing controlled scans in \(d\). This is particularly relevant in regimes where the $d$-dependence of $\Gamma_{\rm collapse}$ differs from that of $\Gamma_{\rm QM}$.

For a cat state \(\ket{\Psi}\) subject only to a thermal Markovian bath (no additional dephasing)—with mechanical energy–relaxation time \(T_1^{m}\) and thermal occupation \(\bar n_{\rm th}\)—the baseline decoherence rate is
\begin{equation}
    \label{eq:thermal_interf}
    \Gamma_{\rm QM}(d)\;\approx\;\frac{(2\bar n_{\rm th}+1)}{2T_1^{m}}
    \left(\frac{d}{X_{\rm zpf}}\right)^{\!2}.
\end{equation}
Although this background rate can in principle be calibrated independently---for instance via temperature sweeps and $T_1^m$ ringdown measurements---the subtraction is only effective when the collapse-induced rate $\Gamma_{\rm collapse}$ constitutes a sizeable fraction of the background $\Gamma_{\rm QM}$.
In the following, we compare the model-dependent collapse rate
\(\Gamma_{\rm collapse}(d;\theta)\) of the CSL and Diósi–Penrose models to this calibrated thermal baseline.

\tocless{\subsubsection}{Expected Diósi–Penrose collapse time\label{sec:diosi_penrose}}

The Diósi–Penrose (DP) model~\citeM{Penrose1996,Diosi1989} ties collapse to gravity by taking, in \refeq{eq:collapse_kernel}, a correlator proportional to the Newtonian potential:
\begin{equation}
    \mathcal{D}_{\rm DP}(\vec r-\vec r')=\frac{4 \pi \,G}{\hbar \lVert \vec r-\vec r'\rVert},
\end{equation}
with \(G\) the gravitational constant. For a Schrödinger-cat state with separation $d$, noting $\vec d$ the displacement vector between the two states of the superposition, we define the mass-density difference field
\begin{equation}
    \delta\mu(\vec r)\;\equiv\;\mu\!\left(\vec r+\tfrac{\vec d}{2}\right)-\mu\!\left(\vec r-\tfrac{\vec d}{2}\right),
\end{equation}
where \(\mu(\vec r)\) is the (classical) mass density. Assuming a rigid-body displacement in either component of the superposition, one can show that the collapse rate is proportional to the gravitational self-energy of \(\delta\mu\):
\begin{equation}
    \begin{aligned}
        \Gamma_{\rm DP}(d) \; & =\; \frac{\Delta E(d)}{\hbar},             \\[2pt]
        \Delta E(d) \;        & =\; 4 \pi G \!\int d^3\vec r\,d^3\vec r'\;
        \frac{\delta\mu(\vec r)\,\delta\mu(\vec r')}{\lVert \vec r-\vec r'\rVert}\,.
        \label{eq:self_energy_diff}
    \end{aligned}
\end{equation}

Because the DP kernel \(\mathcal D_{\rm DP}(\vec r)\) diverges at short distance, one cannot model the mass density \(\mu(\vec r)\) as point-like. The theory does not fix a unique short-distance profile, so the constituent width is often treated as a phenomenological parameter and, ultimately, must be constrained experimentally. To keep the discussion concrete, we follow Diósi’s original formulation~\citeM{Diosi1987} and model the solid as a collection of identical \emph{spherical nuclei} of radius \(a\) (for silicon, \(a\!\approx\!2.7\,\mathrm{fm}\)) rigidly embedded in the object. With this “hard-sphere” nuclear model, \refeq{eq:self_energy_diff} can be evaluated straightforwardly. We provide here the 2 asymptotic behaviors:
\begin{subequations}
    \label{eq:DP_limits}
    \begin{empheq}[left=\Gamma_{\rm DP}(d)\;\approx\;\empheqlbrace]{align}
        &\frac{8 \pi G\,M\,m_{a}}{\hbar\,a} \left(\frac{d}{2a}\right)^{2},&  d\ll a,
        \label{eq:DP_limits_a}\\[6pt]
        &\frac{8 \pi G\,M\,m_{a}}{\hbar\,a} \left(\frac{6}{5}\right),&  d\gtrsim 2a.
        \label{eq:DP_limits_b}
    \end{empheq}
\end{subequations}
where \(M\) is the resonator mass and \(m_{a}\) the nuclear mass. \refeqfull{eq:DP_limits} predicts that the DP-induced decoherence grows quadratically with the separation for small \(d\) and then \emph{saturates} to a value proportional to the total mass once \(d\) exceeds about a nuclear diameter (\(d\gtrsim 2a\)).

Atom and molecule interferometers operate in the large-separation regime (up to \(\sim\!100\) nm for near-field molecular setups and \(\sim\!0.5\) m for atomic interferometers), but with comparatively small masses. Evaluating \(\Gamma_{\rm DP}\) for the largest molecules used so far (\(M\sim 10^{4}\,\mathrm{amu}\)) yields a gravitational decoherence time \(\tau_{\rm DP}=1/\Gamma_{\rm DP}\sim 6\) years even in the saturated regime—well beyond present and foreseeable matter-wave coherence windows (including space-based proposals). This indicates that substantially larger masses would be required to bound DP using molecular interferometry alone.

Conversely, cQAD probes a complementary regime: much larger masses (\(M\gtrsim 10^{15}\,\mathrm{amu}\)) with separations in the attometre–to–femtometre range.
In the limit \(d\ll 2a\), \(\Gamma_{\rm DP}(d)\) and the thermal decay rate \(\Gamma_{\rm QM}(d)\) share the same \(d^{2}\) scaling, making discrimination by \(d\)-scans alone difficult. A sharper, model-specific signature would be to \emph{observe the saturation} of \(\Gamma_{\rm DP}(d)\) as the separation is scanned past \(d\!>\!2a\).

For a consistent comparison with GHz cQAD platforms, we evaluate all systems at the  reference separation \(d_{*}=2a\), which marks the onset of saturation in \refeq{eq:DP_limits}, following Ref.~\citeM{Gely2021}:
for several representative systems we list the phonon number \(|\alpha_{*}|^{2}\) required to reach \(d_{*}\).
For each we then report (i) the DP collapse time \(1/\Gamma_{\rm DP}\)  evaluated in the saturated regime [\refeq{eq:DP_limits_b}], and (ii) the thermal decoherence at the onset, \(1/\Gamma_{\rm QM}(d_*)\).
For GHz devices the situation is unfavorable: in all cases the Diósi–Penrose collapse rate lies at least six orders of magnitude below the thermal decoherence rate, rendering it completely hidden. By contrast, even with our present (non-optimized) MHz resonator the DP rate already approaches the percent level of the thermal baseline, which could be resolvable in a carefully calibrated experiment. Further improvement is realistic: using the soft-clamping technique of Ref.~\citeM{Tsaturyan2017}, our group has demonstrated silicon-nitride membranes with frequencies around $2.5$~MHz and aluminum metallization comparable to the present device, but with $T_{1}^m\!\approx\!2$~s~\citeM{Himanshu_phd}. Incorporating such low-loss oscillators into the platform would move us into a regime where gravitational collapse becomes the dominant decoherence channel (fifth row of the table). Finally, as an illustration of the scaling with frequency, the last row reports parameters for a softly clamped membrane at $1$~MHz, similar to that of Ref.~\citeM{Seis2022}. In this case the predicted DP rate exceeds the thermal decoherence rate by three orders of magnitude, so that gravitational collapse would constitute the primary source of cat-state decay and should be straightforwardly detectable.

Although non-interferometric experiments have already set bounds on DP-type dynamics~\citeM{Donaldi2021, Carlesso2022, Vinante2021}, the present approach would, to our knowledge, be the first \emph{interferometric} test to enter the mass–separation regime where a gravity-related collapse could be directly observed. This distinction matters: non-interferometric tests constrain collapse \emph{indirectly}, from the absence of side-effects (e.g., excess heating or spontaneous radiation). For example, Ref.~\citeM{Donaldi2021} searches, in an underground detector, for spontaneous $\gamma$/X-ray emission generated by collapse-induced stochastic motion of bound charges in solids; interpreting a null result requires assumptions about the electromagnetic coupling of the collapse noise (electrons vs nuclei), the noise spectrum (often taken white up to a high-frequency cutoff), and material-response/background models.
In contrast, interferometric experiments directly test the superposition principle by preparing spatially delocalized states. Although this approach yields a less stringent constraint on the smearing length \(2a\) (nuclear scale here, versus ångström–scale bounds in Ref.~\citeM{Donaldi2021}), it rests on minimal modelling assumptions and probes the core prediction of the collapse model.

\definecolor{lightgreen}{RGB}{181, 245, 219}
\definecolor{chartreuse}{RGB}{100, 217, 169}
\definecolor{lime}{RGB}{22, 201, 128}
\begin{table*}[t]
    \centering
    \begin{tabular}{@{}c
        >{\centering\arraybackslash}m{2.1cm}
        >{\centering\arraybackslash}p{1.55cm}
        >{\centering\arraybackslash}p{1.55cm}
        >{\centering\arraybackslash}p{1.55cm}
        >{\centering\arraybackslash}p{1.55cm}
        >{\centering\arraybackslash}p{1.55cm}
        >{\centering\arraybackslash}p{1.55cm}@{}}
        \toprule
        \textbf{Experiment}                &
        \(\Omega_m/2\pi\)                  &
        \(\mathbf{m}\)                     &
        \(\mathbf{X_{\rm zpf}}\)           &
        \(\mathbf{|\alpha_*|^2}\)          &
        \(\mathbf{T_1^m}\)                 &
        \(1/\Gamma_{\rm{QM}}\)             &
        \(1/\Gamma^{(DP)}_{\rm{collapse}}\)    \\
        \midrule
        HBAR~\citeM{Bild2023}              &
        6~GHz                              &
        16~µg                              &
        0.3~am                             &
        \(8\times 10^7\)                   &
        \(84\,\mu\)s                       &
        0.5~ps                             &
        0.2~$\mu$s                             \\
        \addlinespace[2pt]
        PNC~\citeM{Lee2023}                &
        700~MHz                            &
        8.6~pg                             &
        1~fm                               &
        \(5\)                              &
        4\,$\mu$s                          &
        0.4\,$\mu$s                        &
        0.4\,s                                 \\
        \addlinespace[2pt]
        SAW~\citeM{Satzinger2018}          &
        4~GHz                              &
        0.2~$\mu$g                         &
        3~am                               &
        7$\times 10^5$                     &
        150 ns                             &
        0.1~ps                             &
        20~$\mu$s                              \\
        \addlinespace[2pt]
        \arrayrulecolor{lightgreen}\cline{7-8}
        This work                          &
        4~MHz                              &
        5~ng                               &
        1~fm                               &
        \(7\)                              &
        6~ms                               &
        \multicolumn{1}{|c}{\strut 4\,$\mu$s}
                                           &
        \multicolumn{1}{c|}{\strut 0.6\,ms}
        \\
        \cline{7-8}\arrayrulecolor{black}
        \addlinespace[2pt]
        \arrayrulecolor{chartreuse}\cline{7-8}
        Softly clamped membrane            &
        4~MHz                              &
        5~ng                               &
        1~fm                               &
        \(7\)                              &
        2~s                                &
        \multicolumn{1}{|c}{\strut 2\,ms}  &
        \multicolumn{1}{c|}{\strut 0.6\,ms}    \\
        \cline{7-8}\arrayrulecolor{black}
        \addlinespace[2pt]
        \arrayrulecolor{lime}\cline{7-8}
        1 MHz Softly clamped (prospect)    &
        1~MHz                              &
        80~ng                              &
        0.5~fm                             &
        \(30\)                             &
        130~s                              &
        \multicolumn{1}{|c}{\strut 50\,ms} &
        \multicolumn{1}{c|}{\strut 40\,$\mu$s} \\
        \cline{7-8}\arrayrulecolor{black}
        \addlinespace[2pt]
        \bottomrule
    \end{tabular}
    \caption[Comparison of various mechanical systems for the observation of Diosi-Penrose collapse.]{
        Comparison of various mechanical systems for the observation of Diosi-Penrose collapse:
        The 2 last columns compare the gravitational collapse time predicted by the DP hard sphere model $1/\Gamma_{\rm DP}$ (Eq.~ \eqref{eq:DP_limits_b}) and the thermal decoherence time $1/\Gamma_{\rm{QM}}$ [\refeq{eq:thermal_interf}]. The state size required to reach the separation $d=2a$, expressed in number of phonons $|\alpha_*|^2$, is indicated in the fourth column. Note that the thermal decoherence rate $\Gamma_{\rm QM}$ depends on $|\alpha_*|^2$ via the formula \eqref{eq:thermal_interf}. The systems where $\Gamma_{\rm DP}$ approaches $\Gamma_{\rm QM}$ are highlighted in green. Figure adapted from~\citeM{Gely2021}.
    }
    \label{tab:dp_like}
\end{table*}

\tocless{\subsubsection}{Exclusion bounds on the Continuous Spontaneous Localization (CSL) model}

Several assumptions underlie the Diósi–Penrose proposal—how to coarse-grain the mass density, how strong the putative gravitational noise is, and whether gravity plays any role at all in state-vector reduction. To remain open to alternatives, it is convenient to adopt a theory-agnostic, phenomenological description: Continuous Spontaneous Localization (CSL). Introduced by Pearle and completed in 1990 by Ghirardi, Pearle, and Rimini, CSL assumes a Gaussian correlator without committing to a specific physical origin.
\begin{equation}
    \mathcal{D}_{\rm CSL}(\vec r) = \frac{\lambda}{m_0^2}
    \exp\!\Big(-\tfrac{|\vec r|^{2}}{4 r_C^{2}}\Big)
\end{equation}
with $m_0$ the nucleon mass. The two parameters of the CSL model, $\lambda$ and $r_C$ are respectively the localization rate and spatial correlation length. Since the Kernel $\mathcal{D}_{\rm CSL}(\vec r)$ is finite at short distances, one may treat the solid as a sum of pointlike nuclei with mass density $\mu(\vec r) = \sum_n m_a \delta(\vec r - \vec{r}_n)$, where $m_a$ is the nuclear mass and $\vec r_n$ the position of the $n^{\rm th}$ nucleon in the lattice.
Under these assumptions, the CSL master equation takes the form~\citeM{Bassi2013}:
\begin{equation}
    \label{eq:CSL_full}
    \mathcal{L}_{\rm CSL}[\op \rho]
    = -\,\frac{\lambda}{2\,\pi^{3/2} r_C^{3} m_0^{2}}
    \int d^{3}\!\vec r\,\big[\op m(\vec r),\,[\op m(\vec r),\op \rho]\big],
\end{equation}
with the Gaussian-smeared mass operator
\begin{equation}
    \op m(\vec r)=\sum_{n=1}^{N_{\rm nuc}} m_a \,
    \exp\!\Big(-\tfrac{|\vec r - \vec{\op r}_n|^{2}}{2 r_C^{2}}\Big),
\end{equation}
where $\vec{\op r}_n$ is the position operator of the \(n\)-th nucleus.

Phenomenological studies of CSL commonly consider a sub-micron to micron correlation length, with the canonical benchmark \(r_C = 10^{-7}\,\mathrm{m}\) (GRW) and, more broadly, \(r_C \gtrsim 100~\mathrm{nm}\). In this regime, \(r_C\) exceeds the ångström-scale lattice spacing of solid-state devices by many orders of magnitude, so a coarse-grained continuum mass density is well justified. Moreover, the relevant displacements—zero-point fluctuations \(X_{\rm zpf}\), thermal rms motion, and the cat separations \(d\) considered below—satisfy \(X,d \ll r_C\). A small-displacement expansion then yields~\citeM{Nimmrichter2014}:
\begin{equation}
    \label{eq:CSL_small_d}
    \mathcal{L}_{\rm CSL}[\op \rho] = \;\lambda\,\frac{\alpha_{\rm geo}(r_C)}{r_C^2}\,[\op X,[\op X,\op \rho]],
\end{equation}
where \(\alpha_{\rm geo}(r_C)\) is a geometry factor quantifying how much of the resonator’s mass distribution falls within the localization volume.

Writing the displacement in terms of ladder operators $\op X = X_{\rm zpf} (\op a + \op a^\dagger)$, and discarding counter-rotating terms, \refeq{eq:CSL_small_d} reduces to an
effective phase-insensitive (infinite-temperature) bath:
\begin{subequations}
    \label{eq:CSL_small_d_thermal}
    \begin{align}
        \mathcal{L}_{\rm CSL}[\op\rho]      & = \kappa\big(\sop D[\op a]+\sop D[\op a^\dagger]\big),
        \label{eq:CSL_small_d_thermal_L}                                                                        \\[-2pt]
        \text{with\,\,\,\,\,\,\,\,\,}\kappa & = 2\,\lambda\,\alpha_{\rm geo}(r_C)\,\frac{X_{\rm zpf}^2}{r_C^2},
        \label{eq:CSL_small_d_thermal_kappa}
    \end{align}
\end{subequations}
and \(\sop D[\op L]\op\rho=\op L\op\rho\op L^\dagger-\tfrac12\{\op L^\dagger \op L,\op\rho\}\).
The associated heating (increase of effective temperature) is the standard observable in optomechanical, non-interferometric CSL tests~\citeM{Carlesso2022}.

In a cQED scenario, the cat state of \refeq{eq:cat} subject to the CSL bath of \refeq{eq:CSL_small_d_thermal_L} looses coherence at rate \(2\kappa|\alpha|^{2}\). Using \(d=2|\alpha|X_{\rm zpf}\), we obtain
\begin{equation}
    \label{eq:Gamma_CSL_small_d}
    \Gamma_{\rm CSL}(d)=\lambda\,\alpha_{\rm geo}(r_C)\!\left(\frac{d}{r_C}\right)^{2}\qquad(d\ll r_C).
\end{equation}
In this small separation regime, the CSL dynamics effectively appears as an excess-temperature bath. As a result, $\Gamma_{\rm CSL}$ and the thermal background $\Gamma_{\rm QM}$ have the same $d^2$ scaling, so the CSL contribution can only be isolated via a careful calibration of the thermal background (e.g., temperature dependence). Quantitatively, if $\epsilon$ denotes the fractional uncertainty of the calibrated background, the experiment excludes the CSL parameter region where
\begin{equation}
    \label{eq:csl_criterion}
    \Gamma_{\rm CSL}\!\ge\!\epsilon\,\Gamma_{\rm{QM}}.
\end{equation}

\begin{figure*}[!t]
    \centering
    \includegraphics[width=\textwidth]{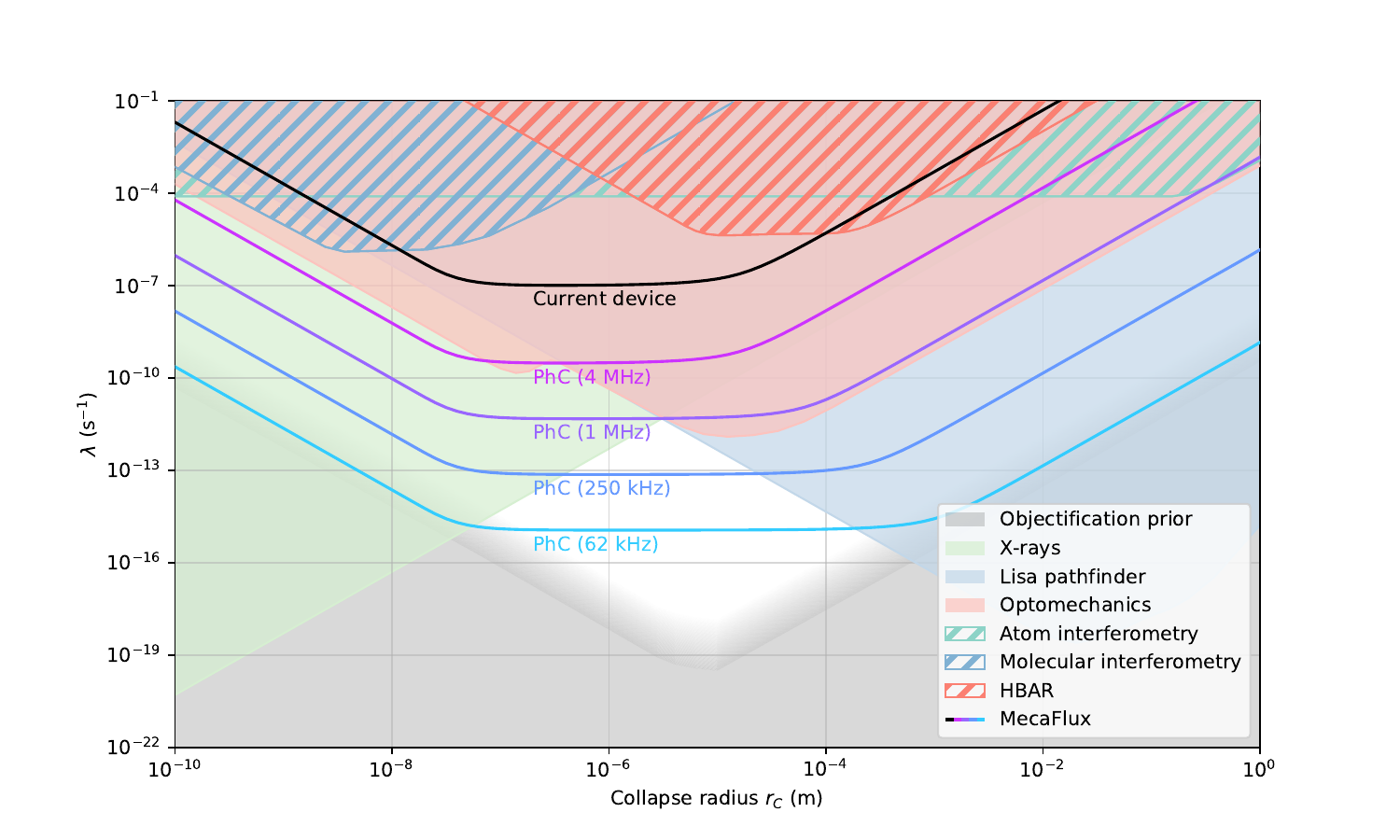}
    \caption[CSL exclusion graph.]{
        \textbf{CSL exclusion graph.}
        In the ($r_C$, $\lambda$) plane, filled areas denote regions excluded by non-interferometric tests, while hatched areas denote interferometric exclusions.
        Filled: the green area derives from spontaneous X-ray emission tests~\citeM{Arnquist2022};
        the blue area from the translation-noise spectrum of LISA Pathfinder~\citeM{Altamura2025}; and the red area from optomechanical measurements of Brownian-noise spectra in cantilevers~\citeM{Vinante2021}.
        Hatched: The cyan region corresponds to atom interferometers (an atomic cloud prepared in a metre-scale spatial superposition)~\citeM{Kovachy2015}, the blue region to molecular interferometry (near-field Talbot–Lau with \(\sim10^{4}\,\mathrm{amu}\) molecules)~\citeM{Eibenberger2013}, and the blue region to a GHz cQAD experiment using a high-overtone bulk acoustic wave resonator prepared in a Schrödinger-cat state~\citeM{Bild2023}.
        The gray band indicates a macro-objectification prior~\citeM{Toros2017}, i.e. a phenomenological requirement that macroscopic superpositions be suppressed on human timescales (not an experimental bound).
        The central white (rhombus-shaped) region remains unconstrained by current measurements, interferometric or otherwise.
        The black line shows the expected reach of the current 4\,MHz device.
        Colored lines indicate the projected reach for soft-clamped phononic-crystal membranes at 4\,MHz, 1\,MHz, 250\,kHz, and 62\,kHz, assuming the empirical \(Q\)-scaling~\citeM{Tsaturyan2017}
        \(Q \simeq 5\times10^{7}\,(2\pi\!\cdot\!4~\mathrm{MHz}/\omega_m)^{2}\).
        All curves mark the locus where \(\Gamma_{\rm CSL}=0.10\,\Gamma_{\rm QM}\) (see main text for modeling details).
    }
    \label{fig:csl_exclusion}
\end{figure*}

\reffigfull{fig:csl_exclusion} summarizes existing constraints and our projected interferometric reach in the \((r_C,\lambda)\) plane.
Our projected device curves are constructed from the criterion~\eqref{eq:csl_criterion} with \(\varepsilon=0.10\),
using the small-separation expression in \refeq{eq:Gamma_CSL_small_d} together with the thermal baseline in \refeq{eq:thermal_interf}.
For the present 4\,MHz device (black line) we use its measured parameters (row~4 of \reftabfull{tab:dp_like}).
For the soft-clamped phononic-crystal membranes (colored lines at 4~MHz, 1~MHz, 250~kHz, and 62~kHz) we assume the commonly observed quality-factor scaling~\citeM{Tsaturyan2017}
\(Q \simeq 5\times10^{7}\,(2\pi\!\cdot\!4~\mathrm{MHz}/\omega_m)^{2}\).
This assumption is realistic: the 4~MHz and 1~MHz PhC membranes (the last two rows of \reftabfull{tab:dp_like}) are close to existing devices~\citeM{Himanshu_phd, Tsaturyan2017}, while the 250~kHz and 62~kHz cases follow the same soft-clamping trendline.
The shape of the projected curves versus \(r_C\) directly reflects the variation of the geometry factor \(\alpha_{\rm geo}, (r_C)\), calculated from the cuboid geometry (formula (S8) of ref.~\citeM{Nimmrichter2014}) with thickness $L_X = 90$~nm, and lateral dimensions $L_Y, L_Z = 50 \,µ\rm{m} \cdot (2 \pi \cdot 4 \,\mathrm{MHz}/\omega_m)$: a left "knee" when \(r_C\) drops below the thickness \(L_X\), a plateau for \(L_X\!\lesssim\!r_C\!\lesssim\!L_Y,L_Z\), and a right "knee" as \(r_C\) exceeds the lateral size.
Similarly, for the
HBAR interferometric bound, we use the device parameters reported in Ref.~\citeM{Bild2023} and evaluate \(\alpha_{\rm geo}(r_C)\) for a cylindrical (disk) geometry (formula (S9) of ref.~\citeM{Nimmrichter2014}, with radius $13.5$~µm, and thickness $435$~µm).

The strongest interferometric bounds to date are shown as hatched overlays.
Interference of atomic clouds prepared in metre-scale spatial superpositions~\citeM{Kovachy2015} constrains
\(\lambda \gtrsim 10^{-4}\,\mathrm{s}^{-1}\) at very large \(r_C\) (cyan hatches).
Molecular interferometry~\citeM{Eibenberger2013} using near-field Talbot–Lau fringes of \(\sim 10^{4}\,\mathrm{amu}\) molecules
sets bounds \(\lambda \gtrsim 10^{-6}\,\mathrm{s}^{-1}\) as \(r_C\) approaches the achieved delocalization length (tens to \(\sim 100\) nm; blue hatches).
Finally, high-overtone bulk acoustic wave resonators (HBAR) prepared in Schrödinger-cat phononic states with GHz cQAD techniques~\citeM{Bild2023} constrain CSL for \(r_C\) between the resonator’s lateral dimension (\(\sim 10~\mu\mathrm{m}\)) and its thickness (\(\sim 0.5~\mathrm{mm}\)), although the exclusion depth is limited to \(\lambda \approx 5\times 10^{-6}\,\mathrm{s}^{-1}\) by the small zero-point amplitude and finite lifetime of such GHz devices (red hatches).
By contrast, current non-interferometric exclusions, shown as filled areas, cover a wider portion of the \((\lambda, r_C)\) plane.
Spontaneous X-ray emission tests in underground facilities~\citeM{Arnquist2022} provide the strongest bounds below \(r_C \approx 1~\mu\mathrm{m}\) (green region).
Optomechanical cantilever measurements of Brownian spectra~\citeM{Vinante2021} constrain \(\lambda \lesssim 10^{-12}\,\mathrm{s}^{-1}\) for intermediate \(r_C \sim 1\text{–}100~\mu\mathrm{m}\) (red region).
The translation-noise spectrum of LISA Pathfinder~\citeM{Altamura2025} yields the tightest constraints at large \(r_C\) (blue region).
We also indicate a phenomenological macro-objectification prior~\citeM{Toros2017}—parameter choices that would fail to suppress macroscopic superpositions on human time scales (light-gray band; not an experimental bound).
The central white region highlights parameter space that remains unconstrained by any existing measurement.

The MHz cQAD platform developed in this work provides a decisive interferometric reach across the range \(L_X \lesssim r_C\!\lesssim \!L_Y, L_Z\).
For the present (non-optimized) 4\,MHz device (black curve), the projected exclusion attains \(\lambda \sim 10^{-7}\,\mathrm{s}^{-1}\) for correlation lengths \(r_C \sim 100\)~nm--100~µm, exceeding existing interferometric bounds over that interval.
Phononic-crystal membranes, owing to their exceptionally high mechanical quality factors extend well beyond existing interferometric exclusions.
The coloured curves illustrate the frequency trend, with sub-MHz designs extending well beyond existing interferometric exclusions and entering the previously unconstrained (white) region of the \((r_C,\lambda)\) plane.
Beyond enlarging the excluded region, these measurements provide the most direct tests of the quantum superposition principle in this parameter range: they measure the coherence of genuine mesoscopic superpositions, rather than inferring collapse from \emph{indirect} observables (e.g., excess heating or spontaneous radiation) that depend on model-specific assumptions about the collapse-noise spectrum, cutoffs, and dissipation pathways~\citeM{Carlesso2022}. Consequently, the resulting limits yield sharper, model-specific constraints on \((r_C,\lambda)\) than non-interferometric approaches.

\par\bigskip

\tocless{\subsubsection}{Conclusion}
Advancing cQAD into the MHz regime with long \(T_1^{m}\) and large \(X_{\rm zpf}\) enables \emph{interferometric} tests of objective-collapse dynamics.
For the Diósi–Penrose (DP) model, \emph{nuclear-scale} spatial separations become directly accessible, offering a strong, model-specific experimental signature.
For Continuous Spontaneous Localization (CSL), the experiment probes the \emph{small-separation} regime \(d\!\ll\! r_{C}\), where the collapse field manifests as an effectively phase-insensitive excess-temperature bath.
Importantly, these measurements would provide the first tests using genuine quantum superpositions entering the relevant parameter range.

\begin{acknowledgments*}
    We thank Luis Najera for early experimental developments.
    We thank Emilio Rui and Philippe Campagne-Ibarcq for insightful discussions. We thank Jean-Michel Raimond for carefully reading the manuscript.
    We thank the SQC team at Institut Néel for providing a TWPA.
    This work was supported by the Agence Nationale de la Recherche under projects \textsc{MecaFlux} (ANR-21-CE47-0011), \textsc{Ferbo} (ANR-23-CE47-0004), and RobustSuperQ (ANR-22-PETQ-0003).
    This work has been supported by Region Île-de-France in the framework of DIM \textsc{Sirteq} (project CryoParis) and DIM QuanTiP (project COCONUT), and by Sorbonne Université through the \textsc{HyQuTech} “Emergence” program. K.~G. acknowledges support from the Quantum Information Center Sorbonne (QICS doctoral fellowship), H.\,P. is funded by the CNRS–University of Arizona joint Ph.D. program.
\end{acknowledgments*}

\tocless{\section*}{Author contributions}
K.G., L.B. and R.R. jointly designed and simulated the qubit sample, L.B., K.G. and P.M. micro-fabricated the sample. K.G., S.D. and R.R. performed measurements, K.G. and R.R. performed data analysis, R.R. and E.F. developed the theoretical framework for sideband asymmetry. R.R. derived the bounds on collapse model parameters under the supervision of A.T.; H.P. and T.J. developed the flip-chip process and fabricated the membrane. E.F., Z.L. and W.C.S. provided essential guidance on the experimental implementation, contributed key conceptual ideas, and critically read and revised the manuscript.
K.G., R.R., T.J. and S.D. contributed equally to writing the article.
All authors participated in discussions, reviewed the manuscript, and approved its final version.
S.D and T.J. supervised the project.

\tocless{\section*}{Competing interests}

The authors declare no competing interests.

\tocless{\section*}{Data availability}
The data that support the findings of this study are available from the corresponding authors upon reasonable request.

\bibliographystyleM{naturemag}

{\let\oldaddcontentsline\addcontentsline
    \renewcommand{\addcontentsline}[3]{}\bibliographyM{bib-methods}
    \let\addcontentsline\oldaddcontentsline
}

\renewcommand{\thefigure}{S\arabic{figure}}
\renewcommand{\thetable}{S\arabic{table}}
\renewcommand{\theequation}{S\arabic{equation}}
\renewcommand{\thesection}{S\arabic{section}}
\renewcommand{\thesubsection}{\thesection.\arabic{subsection}}

\makeatletter
\renewcommand{\p@subsection}{}
\makeatother

\clearpage
\onecolumngrid

\begin{center}
    {\LARGE\bfseries Supplementary Information}\\[0.5em]
    {\large “Probing the quantum motion of a macroscopic mechanical resonator with a radio-frequency superconducting qubit”}\\[2em]
\end{center}

\twocolumngrid
\tableofcontents

\section{Design and fabrication}

\tocless{\subsection}{Qubit design and fabrication}
The fluxonium qubit is fabricated on a 280~µm intrinsic Si wafer. All superconducting wiring (except for junctions) is defined from a sputtered 200~nm Nb film by wafer‐scale UV‐laser lithography (S1805 resist) and dry etching. Aluminum spacers (600~nm tall) set the nominal flip‐chip gap and are patterned via a bilayer resist (LOR20B/AZ~5214E), UV lithography, Al evaporation, and lift‐off. Josephson junctions are formed using a Dolan‐bridge bilayer (MMA~EL13/PMMA~A3) and double‐angle Al evaporation ($\pm 22^\circ$) followed by NMP liftoff.
After fabrication, the wafer is diced, and the chips are wire-bounded to a printed‐circuit board. Qubit control is achieved via a capacitively coupled waveguide (cyan in \reffig[(b)]{fig:fig_design}); readout employs a $\lambda/4$ resonator (5.59~GHz, light pink in \reffig[(b)]{fig:fig_design}) in series with an inductively coupled $\lambda/4$ Purcell filter (5.92~GHz); flux biasing uses an on‐chip flux line (brown in \reffig[(b)]{fig:fig_design}) for fine tuning near \(\varphi_{\rm ext}=\pi\) plus an external coil for coarse adjustments.

\begin{figure}
    \centering
    \includegraphics[width=1\linewidth]{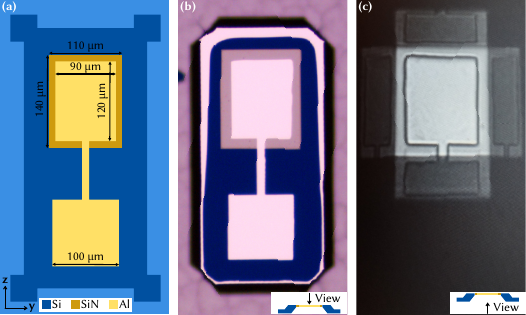}
    \caption{(a) Top: suspended SiN rectangular membrane in orange (only the non metalized edge is visible).
        Aluminum layer in gold.
        Silicon substrate in blue.
        Bottom: square biasing electrode lying on the substrate.
        (b) Optical microscope image of the front-side of the final device.
        The suspended membrane and the biasing electrode sit on a silicon mesa. The typical rough etch bottom, arising from KOH etch can be seen at the edges of the picture.
        (c) Optical image of the same sample, seen from the back side.
        The released metalized SiN membrane can be seen at the top.
        Mirror images on the four edges of the membrane come from the smooth crystalline planes obtained after etching with KOH trough the substrate.}
    \label{fig:design}
\end{figure}
\tocless{\subsection}{Membrane design and fabrication}
The mechanical resonator is fabricated from a 350~µm-thick (100) Si wafer coated on both sides with 100~nm of high-stress SiN (\(\gtrsim800\)\,MPa). On the backside, a rectangular opening is defined in the SiN by UV lithography (AZ~5214E) and RIE, and the silicon is then anisotropically etched in 30~\% KOH at 85 °C until the membrane is released.
On the front side, SiN is removed by UV lithography (AZ~5214E) and RIE everywhere except around the membrane and bias-pad. A second KOH etch (same chemistry, timed to remove $\approx$25~µm of silicon) recesses the substrate everywhere except in this zone, reducing the area of close proximity between the two chips and mitigating the risk of a dust particle compromising the flip-chip process.
The wafer is then immersed in 10~\% HF for 5 min to remove surface oxide and uniformly thin down the released SiN to 90~nm. Finally, UV lithography and evaporation of 30~nm Al (followed by lift-off) define the central membrane electrode and the square bias pad (\reffig[]{fig:design}). Room-temperature Michelson interferometry (prior to metallization) reveals the fundamental mode at 2.54~MHz, the first asymmetric mode at 3.75~MHz, and the second asymmetric mode, which is used in this work, at 4.30~MHz. For this mode, we observed a room-temperature quality factor before metallization of \(Q\approx6\times10^4\).

\tocless{\subsection}{Flip‐Chip Assembly}
The flip-chip assembly is performed using an \textit{MJB4} mask aligner for micropositioning and a custom-built holder.
The chips are aligned with an estimated accuracy of 5\um and brought into contact.
Bonding is achieved using drops of \textit{Dymax OP-67-LC} UV-curing glue at the corners of the membrane chip.
The aluminum spacers on the qubit chip nominally set the inter-chip gap to 600 nm.
While room-temperature Fizeau interferometry on unmetalized test samples confirmed this spacing, cryogenic qubit spectroscopy consistently reveals a gap of 1–3 µm, which we attribute to differential bending under thermo–mechanical stress.

\section{Hamiltonian derivation}
\label{sec:qubit-membrane_coupling}

\subsection{Effective mass}
\label{sec:effective_mass}
In this section, we derive the effective mass of the vibrational mode.
The membrane is a rectangular suspended film of width $L_y = 110~\mathum$, length $L_z = 140~\mathum$, and thickness $t = 90~\mathnm$.
Including the added mass of the $30~\mathnm$-thick aluminum electrode with area $90~\mathum\times120~\mathum$, the physical mass $M = 5.3~\mathrm{ng}$ is obtained using the mass densities of silicon nitride $\rho_\text{SiN}=3200~\mathrm{kg}/\mathrm{m}^3$ and aluminum $\rho_\text{Al}=2700~\mathrm{kg}/\mathrm{m}^3$.
The membrane supports a family of out-of-plane vibrational modes, indexed by the number of antinodes $(m,n)$ in the lateral and longitudinal directions, with the displacement profile $u_{m, n}(y, z)$ given by
\begin{equation}
    \label{eq:mode_profile}
    \lambda \epsilon(y, z)
    \sin\left(m \pi  \frac{y\! -\! L_y/2}{L_y}\right)
    \sin\left(n \pi \frac{z\!-\! L_z/2}{L_z}\right).
\end{equation}
Here, $\epsilon(y,z)$ is a regularization function which, for thin films under high tensile stress $\sigma_\text{SiN}$, equals unity everywhere except in an exponentially small region near the clamping edges~\citeS{Yu2012} $y=\pm L_y/2$ and $z=\pm L_z/2$. While this factor is crucial to capture mechanical damping in the membrane~\citeS{Yu2012}, it has negligible effect on the motional mass and will henceforth be set to unity.
As the mode profiles \eqref{eq:mode_profile} are solutions of the linear elasticity equations, there is a freedom in choosing the normalization constant $\lambda$, leading to an ambiguity in the motional mass $m_\lambda$.
The kinetic energy of a single (2, 1)-mode with amplitude $X$ is given by
\begin{equation}
    \label{eq:membrane_kinetic_energy}
    T = \frac{1}{2}\dot{X}^2
    \iint \rho_{\text{s}}\,u_{2,1}^2(y,z)\,\mathrm{d}y\,\mathrm{d}z,
\end{equation}
where $\rho_{\text{s}}$ is the surface mass density of the membrane.
Identifying to the harmonic oscillator kinetic energy  $T = \frac{1}{2} m_\lambda \dot{X}^2$ leads to the effective mass
\begin{equation}
    \label{eq:m_lambda}
    m_\lambda = \iint \rho_{\text{s}}\,u_{2, 1}^2(y,z)\,\mathrm{d}y\,\mathrm{d}z
    = M\,\frac{\lambda^2}{4}.
\end{equation}
While any normalization factor $\lambda$ is admissible provided the effective mass is redefined according to
\refeq{eq:m_lambda}, a natural choice, given the electrostatic coupling scheme employed in this work, is to choose $\lambda$ such that
\begin{equation}
    \iint_\mathcal{A} u_{2,1}(y,z)\,\mathrm{d}y\,\mathrm{d}z = \mathcal{A},
\end{equation}
where $\mathcal{A}$ denotes the area of overlap in each of the capacitors $C_m^\pm$. This choice validates the parallel-plate approximation for small displacements $X$, yielding
\begin{equation}
    C_m^\pm(X) \simeq C_m\bigl(1 \mp X/d\bigr).
\end{equation}
Indeed,
\begin{equation}
    \label{eq:C_m_expansion_integral}
    \begin{aligned}
        C_m^\pm(X)
         & = \iint_\mathcal{A}
        \frac{\epsilon_0}{\,d \pm X\,u_{2,1}(y,z)\,}\,\mathrm{d}y\,\mathrm{d}z \\
         & \approx \frac{\epsilon_0\mathcal{A}}{d}
        \left(1 \mp \frac{X}{d\mathcal{A}}\iint_\mathcal{A}u_{2,1}(y,z)\,\mathrm{d}y\,\mathrm{d}z\right).
    \end{aligned}
\end{equation}
With this normalization one finds $\lambda=1.3$ and hence
\[
    m_\lambda = M\,\frac{\lambda^2}{4} = 2.3~\mathrm{ng}.
\]
This effective mass corresponds to the mass of the vibrational mode as seen by the fluxonium electrodes.
The second-order term in the expansion \eqref{eq:C_m_expansion_integral} differs from the parallel-plate approximation by $5.5\%$ and will be neglected in further analytical calculations.

\subsection{Circuit quantization}
\label{sec:circuit_quantization}
\begin{figure}
    \centering
    \includegraphics[width=1\linewidth]{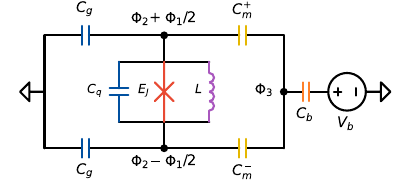}
    \caption{
        \textbf{Equivalent electrical circuit of the flip-chip device.}
        Schematic diagram showing the equivalent circuit and defining the component symbols and node labels used throughout the text.
    }
    \label{fig:circuit_notations}
\end{figure}

Qubit–membrane coupling is achieved via the two capacitors~\citeS{Viennot2018} $C_m^+$ and $C_m^-$, formed between the qubit’s conducting pad and the metallized membrane (illustrated in \reffig[(c)]{fig:design}).
The fluxonium qubit consists of a floating superconducting loop that is symmetrically coupled to ground through two identical capacitors \(C_g\).
The electrical schematic is shown in \reffig{fig:circuit_notations}, where any additional parasitic capacitances have been neglected.
The dynamics are described by the membrane displacement coordinate $X$ and the vector of flux-node coordinates
\begin{equation}
    \vec{\Phi}^T = \bigl(\Phi_1,\;\Phi_2,\;\Phi_3\bigr),
\end{equation}
where $\Phi_1$ is the flux across the qubit, and $\Phi_{2,3}$ denote the fluxes at the coupling nodes (see \reffig{fig:circuit_notations}).

We express the total Lagrangian as the sum of mechanical and electrical contributions:
\begin{equation}\label{eq:L_total}
    \mathcal{L}
    = \mathcal{L}_\mathrm{mec}(X,\dot X)
    + \mathcal{L}_\mathrm{elec}(\Phi_1,\dot{\vec\Phi},X)\,,
\end{equation}
where overdots indicate time derivatives.
The membrane Lagrangian is that of a harmonic oscillator:
\begin{equation}
    \label{eq:L_mec}
    \mathcal{L}_\mathrm{mec}
    = \frac12\,m\,\dot X^2 - \frac12\,m\,\omega_m^2\,X^2,
\end{equation}
with $m$ the effective mass of the membrane and $\omega_m$ its resonance frequency.
The electrical Lagrangian is split into kinetic and potential parts, $\mathcal{L}_\mathrm{elec}=T_\mathrm{elec}-V_\mathrm{elec}$:
\begin{align}
    T_\mathrm{elec}
     & = \frac{C_{g}}{2}\Bigl(\dot\Phi_2 - \tfrac12\dot\Phi_1\Bigr)^2
    + \frac{C_{g}}{2}\Bigl(\dot\Phi_2 + \tfrac12\dot\Phi_1\Bigr)^2
    \nonumber                                                         \\
     & \quad
    + \frac{C_m^+}{2}\Bigl(\dot\Phi_3 - \dot\Phi_2 - \tfrac12\dot\Phi_1\Bigr)^2
    + \frac{C_m^-}{2}\Bigl(\dot\Phi_3 - \dot\Phi_2 + \tfrac12\dot\Phi_1\Bigr)^2
    \nonumber                                                         \\
     & \quad
    + \frac{C_b}{2}\bigl(V_b - \dot\Phi_3\bigr)^2
    + \frac{C_q}{2}\,\dot\Phi_1^2,
    \label{eq:T_elec}
    \\[6pt]
    V_\mathrm{elec}
     & = -E_J\cos\bigl(2\pi\Phi_q/\Phi_0\bigr)
    + \frac{\Phi_1^2}{2L}.
    \label{eq:V_elec}
\end{align}
Here, $C_g$ is the ground capacitance, $C_b$ the membrane bias capacitance, $C_q$ the qubit shunt capacitance, $V_b$ the bias voltage, $E_J$ the Josephson energy, $\Phi_0=h/2e$ the flux quantum, and $L$ the inductance of the fluxonium.

For the Legendre transform we rewrite $T_\mathrm{elec}$ in matrix form:
\begin{equation}\label{eq:T_matrix}
    T
    = \frac12\sum_{i,j=1}^3\dot{\Phi}_i\,M_{ij}(X)\,\dot{\Phi}_j
    - C_b\,V_b\,\dot\Phi_3,
\end{equation}
with the position‐dependent capacitance matrix
\[
    M(x) =
    \begin{pmatrix}
        \tfrac12\,(C_g + C_m^\Sigma + 2C_q) & C_m^\Delta            & -C_m^\Delta         \\[4pt]
        C_m^\Delta                          & 2\,(C_g + C_m^\Sigma) & -2\,C_m^\Sigma      \\[4pt]
        -C_m^\Delta                         & -2\,C_m^\Sigma        & C_b + 2\,C_m^\Sigma
    \end{pmatrix},
\]
with $C_m^\Sigma \equiv (C_m^+ + C_m^-)/2$ and $C_m^\Delta \equiv (C_m^+ - C_m^-)/2$ are symmetric and antisymmetric combinations of the membrane-qubit capacitances.
The canonical flux‐momenta are defined by $Q_k = \partial T/\partial \dot{\Phi}_k$, where $k=1,2,3$.
This yields
\begin{equation}
    \dot{\Phi}_k = \sum M_{ki}^{-1}\,\widetilde{Q}_i\,,
    \text{ where }
    \widetilde{Q}_i =  Q_i + C_b\, V_b\, \delta_{i3}
\end{equation}
Here $\delta_{ij}$ is the Kronecker delta.
The membrane momentum is simply $P_X = m\,\dot X$.
Performing the Legendre transform, $H = \sum_{i=1}^{3}Q_i\,\dot\Phi_i + P_X\,\dot X - \mathcal{L}$, yields the Hamiltonian
\begin{align}
    \label{eq:H_full}
    H & = \frac12 \sum_{i,j=1}^3\widetilde{Q}_i\,M^{-1}_{ij}\,\widetilde{Q}_j
    - E_J\cos\bigl(2\pi\Phi_1/\Phi_0\bigr)
    + \frac{\Phi_1^2}{2L}
    \nonumber                                                                 \\
      & \quad
    + \frac{P_X^2}{2m}
    + \frac12\,m\,\omega_m^2\,X^2.
\end{align}
Since $\mathcal{L}$ [\refeq{eq:L_total}] does not depend on $\Phi_2$ or $\Phi_3$, these coordinates are cyclic and their conjugate charges are strictly conserved by Noether’s theorem.
Physically, these conserved quantities correspond to the static charges on the fluxonium and membrane islands, and may therefore be treated as fixed scalar parameters in the subsequent analysis.

We quantize the system by introducing the canonical variables for the qubit phase $\varphi$ and Cooper pair number $n$, along with the dimensionless canonical variables for the membrane position $x$, momentum $p$ and $\aop$ ($\aopd$) the mechanical annihilation (creation) operators:
\begin{equation}
    \begin{alignedat}{4}
        \op{\varphi} & = 2\pi \frac{\op{\Phi}_1}{\Phi_0},    & \qquad
        \op{x}       & = \frac{\op{X}}{2\, X_{\mathrm{zpf}}} & = \frac{\aop + \aopd}{2},
        \\[6pt]
        \op{n}       & = \frac{\op{Q}_1}{2 e},               & \qquad
        \op{p}       & = \frac{\op{P}}{2 P_{\mathrm{zpf}}}   & = i\frac{\aopd - \aop}{2},
    \end{alignedat}
\end{equation}
where $X_\mathrm{zpf} = \sqrt{\hbar / (2 m \omega_m)}$ and $P_\mathrm{zpf} = \sqrt{\hbar m \omega_m / 2}$ are the membrane zero-point fluctuations of position and momentum, respectively.
These variables satisfy the commutation relations $[\op{\varphi}, \op{n}] = i$ and $[\op{x}, \op{p}] = i/2$, and operators from different sets commute.
The capacitance matrix depends on the membrane position, $\op{M}^{-1}_{ij} = \op{M}^{-1}_{ij}(\op{x})$, and must therefore be treated as an operator.
In this basis the Hamiltonian assumes the familiar form for a fluxonium coupled to a harmonic mode:
\begin{align}
    \op{H}
     & = 4\,\op{E}_C(\op{x})\bigl[\op{n} + \op{n}_g(\op{x})\bigr]^2
    - E_J\cos\op{\varphi}_q
    + \tfrac12\,E_L\,\op{\varphi}_q^2
    \nonumber                                                       \\[6pt]
     & \quad
    + \hbar\omega_m \bigl(\op{p}^2 + \op{x}^2\bigr)
    + \op{F}_m(\op{x}),
\end{align}
where
\begin{gather}
    E_L = \left(\frac{\Phi_0}{2\pi}\right)^2 \frac{1}{L},
    \quad
    \op{E}_C(\op{x}) = \frac{e^2}{2}\,\op{M}^{-1}_{11}(\op{x}),
    \nonumber \\
    \op{n}_g(\op{x}) = \frac{e}{4\,\op{E}_C(\op{x})}
    \sum_{i=2}^3\widetilde Q_i\,\op{M}^{-1}_{1i}(\op{x}).
    \nonumber
\end{gather}
This form of the Hamiltonian shows that the coupling between the fluxonium and the membrane induces an additional charge $\op{n}_g(\op{x})$ on the fluxonium.
The membrane also experiences an additional electrostatic potential $\Hms(\op{x})$, which, as will be shown later, leads to a reduction of the membrane frequency.
The general form of $\Hms$ is given by
\begin{equation}
    \Hms = 4 \op{E}_C \op{n}_g^2 + \op{F}_m
    = \frac{1}{2}
    \sum_{i,j=2}^3
    \widetilde{Q}_i\,
    \op{M}^{-1}_{ij}\,
    \widetilde{Q}_j.
\end{equation}

\subsection{Qubit–membrane coupling}
\label{sec:qubit_membrane_coupling}

For small membrane displacements we perform a first‐order expansion of the coupling Hamiltonian in the dimensionless coordinate $\op{x}$, yielding
\begin{equation}
    \begin{aligned}
        H_{\mathrm{coupling}}
         & = 8\,\op{E}_C(\op{x})\,\op{n}_g(\op{x}) \\
         & \;\approx\;
        4E_C\,
        \frac{\partial C_m^+}{\partial x}\,
        \frac{V_\text{eff}(V_b)}{e}\,
        \op{n}\,\op{x}\,,
        \label{eq:H_coupling}
    \end{aligned}
\end{equation}
Here, $E_C = e^2/(C_g + C_m + 2 C_q)$ is the unperturbed capacitive energy of the fluxonium (neglecting all terms beyond linear order in $x$), and
\[
    \frac{\partial C_m^\pm}{\partial x}
    = 2\,X_{\mathrm{zpf}}\,\frac{\partial C_m^\pm}{\partial X}
\]
relates the derivative with respect to the dimensionless coordinate $\op{x}$ back to the physical membrane displacement $X$.
The effective bias, $\Veff(V_b)$, is given by:
\begin{equation}
    \label{eq:coupling-bias-coeff}
    \begin{aligned}
        \Veff(V_b)
         & = \frac{\,C_g\,V_b - Q_2/2 + \,(C_g/C_b)\,Q_3}
        {C_g + C_m + 2\,(C_g\,C_m/C_b)}                   \\[6pt]
         & \;\longrightarrow\;
        \frac{C_g\,V_b - Q_2/2}{C_g + C_m}\,,
    \end{aligned}
\end{equation}
where the last expression indicates the limit $C_g/C_b\!\ll\!1$.
This indicates that trapped charges, $Q_2$ and $Q_3$, act as an effective bias on the qubit.
Introducing the dilution factor $\beta = C_g/(C_g + C_m)$ and the offset voltage $V_\text{offset} = Q_2/2 C_g$, we have
\begin{equation}
    V_\text{eff}(V_b) = \beta (V_b - V_\text{offset})
\end{equation}

Projecting onto the lowest two levels of the fluxonium and invoking the rotating‐wave approximation yields the Jaynes–Cummings Hamiltonian,
\begin{equation}\label{eq:coupling-JC}
    \op{H}_{JC}/\hbar
    = \omega_m\,\op{a}^\dagger\op{a}
    + \frac{\omega_q}{2}\,\op{\sigma}_z
    + i\,\frac{\Omega}{2}\,
    \bigl(\op{a}^\dagger\,\op{\sigma}_- - \op{\sigma}_+\,\op{a}\bigr),
\end{equation}
where the effective vacuum Rabi frequency is
\begin{equation}
    \Omega
    = \omega_q\,\bigl|\braket{g|\op{\varphi}|e}\bigr|\,
    \frac{d C_m^+}{d x}\,\frac{\Veff}{2e}\,.
\end{equation}
Here, $\omega_q$ is the qubit transition frequency and $\op{\sigma}_\pm$ are the qubit raising and lowering operators; the particular phase convention for $\op{\varphi}_q$ has been chosen to simplify later expressions.

Using the independently determined parameters $\omega_q/2\pi=2.35\ \mathrm{MHz}$ (Ramsey experiment), $|\braket{g|\op{\phi}|e}|=3.04$ (numerical evaluation using parameters from the qubit spectroscopy), $V_\mathrm{eff}=\beta\,(V_b-V_\mathrm{offset})=5.6\ \mathrm{V}$ (Ansys simulation for $\beta$), and ${dC_m^+}/{dx}=2X_{\mathrm{zpf}}{C_m}/{d}$, together with $C_m(d=2.5\,\upmu\mathrm{m})=13.9\,\mathrm{fF}$ (Ansys simulation) and $X_{\mathrm{zpf}}=0.9\,\mathrm{fm}$, we obtain
\begin{equation}
    \Omega^\text{th}/2 \pi = 1.31(33)\,\text{kHz},
\end{equation}
A value that is in good qualitative agreement with the value $\Omega/2\pi = 1.50(8)$~kHz extracted directly from the signal-to-noise ratio of the spectrum analyzer experiment (see \refeq{eq:s_mm_final_spectrum}).

\subsection{Electrostatic spring softening}
\label{sec:membrane_bias}

A bias voltage $V_b$ on the membrane produces an electrostatic potential $\op{H}_{ms}$ that effectively softens the mechanical spring and lowers the resonant frequency.
To capture this effect, we expand $\Hms$ to second order in the dimensionless displacement $\op{x}$, express it in terms of the ladder operators and retain only energy‐conserving terms.
The membrane Hamiltonian then becomes
\begin{equation}
    \begin{aligned}
        \op{H}_{\mathrm{membrane}}
         & = \hbar\,\omega_m\bigl(\aopd\,\aop + \tfrac12\bigr) \;+\; \op{H}_{ms}               \\[4pt]
         & = \hbar\,\bigl(\omega_m + \delta\omega_m\bigr)\bigl(\aopd\,\aop + \tfrac12\bigr)\,,
    \end{aligned}
\end{equation}
where the frequency shift is
\begin{equation}
    \delta\omega_m
    = - \zeta (V_b - V_\text{offset})^2\label{eq:delta_omega_m}
\end{equation}
with
\begin{equation}
    \zeta =
    2\,C_m\,(C_g + 2C_q)\,
    \frac{E_C}{\hbar e^2}\,
    \frac{X_{\mathrm{zpf}}^2}{d^2}\,\beta^2.
    \label{eq:zeta}
\end{equation}
Here we have used the parallel‐plate model $C_m^\pm(X) = {C_m}/(1 \pm X/d)$ to calculate the derivatives ${d C_m^\pm}/{d X}$ and ${d^2 C_m^\pm}/{d X^2}$.

The electrostatic force thus acts as a “negative spring”, lowering the membrane’s resonance frequency.
In \reffig[(a)]{fig:interaction} we plot this softening and extract a tuning coefficient
\[
    \zeta/2\pi = -0.980\,\mathrm{Hz}/\mathrm{V}^2.
\]
By combining \refeq{eq:zeta} with the capacitance \(C_m(d)\) extracted from our ANSYS Q3D simulations and the gap \(d\), independently determined (with its uncertainty) in \refsec{sec:qubit-membrane_distance}, we obtain
\[
    \zeta_{\rm th}/2\pi = -1.14(42)\,\mathrm{Hz}/\mathrm{V}^2.
\]
Because $\zeta \propto d^{-3}$, this parameter-free agreement between \(\zeta\) and \(\zeta_{\rm th}\) provides a stringent consistency check on our membrane–qubit separation.

\subsection{Offset voltage discussion}
\label{sec:offset-voltage-discussion}

In our measurements, we consistently observed a residual static charge on the membrane, which offsets the voltage bias required to cancel the electrostatic force by an amount \(V_{\mathrm{offset}}\) (see Fig~2a).
This offset arises from trapped charges \(Q_2\) and \(Q_3\) on the islands between the capacitors (see \refeq{eq:coupling-bias-coeff}).
Although the qubit remains operational for a DC-bias voltage up to \(|V_b| \lesssim 9\,\mathrm{V}\), the effective coupling strength depends on the difference \(V_b - V_{\mathrm{offset}}\).
Therefore, the presence of this offset can be used to increase the coupling strength.

We found that \(V_{\mathrm{offset}}\) slowly drifts toward the applied bias over a timescale of weeks.
To restore the maximal coupling range, we deliberately “reset” the offset by applying a large negative bias of \(V_b = -20\,\mathrm{V}\) to the membrane for several days.
While this high voltage temporarily disables measurements, it reliably shifts \(V_{\mathrm{offset}}\) closer to \(-9\,\mathrm{V}\).
After this procedure, we verify the new offset value using Ramsey-type membrane spectroscopy (see \reffig[(a)]{fig:interaction}) and ensure that it lies within the desired measurement range.

\subsection{Qubit-membrane distance estimation}
\label{sec:qubit-membrane_distance}

\begin{figure}
    \includegraphics{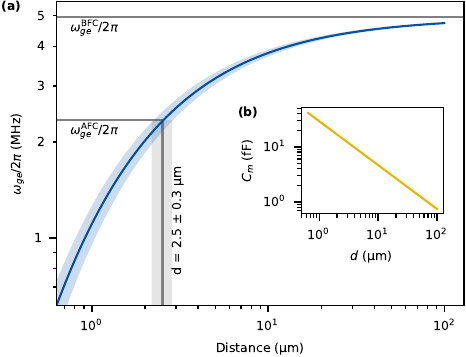}
    \caption{
        \textbf{Distance estimation.}
        Empirical approximation of the capacitance found in Ansys Q3D simulations $C_m(d) = \epsilon_0\times 54 \times 10^{-9} / d^{\,0.8}$, where both $\epsilon_0$ and $d$ are in SI units.
    }
    \label{fig:distance-estimation}
\end{figure}

The value of the vacuum‐gap capacitors $C_m^\pm$ depends sensitively on the separation distance $d$ between the fluxonium and the membrane.
To estimate $d$ accurately, we measured the same fluxonium device both before and after the flip‐chip assembly.
Plasmon transitions ($\omega_{\text{ef}}, \omega_{\text{gh}} \propto E_C^{1/2}$) are only weakly dependent on the capacitive energy and are significantly shifted by coupling to nearby array modes.
In contrast, the low‐frequency qubit transition $\omega_q$ scales exponentially with $E_C$, making it a more precise probe of the capacitor gap.
At half a flux quantum bias, $\omega_q$ follows the approximation~\citeS{Vakhtel2024, Ardati2024}:
\begin{equation}
    \omega_{ge} =
    \frac{8\cdot 2^{\frac34}}{\sqrt{\pi}}
    \,E_J^{\frac34}\,E_C^{\frac14}(d)\,
    \exp\!\Bigl[-\sqrt{\tfrac{8E_J}{E_C(d)}} + \tfrac{14\zeta(3)\,E_L}{\sqrt{8E_J E_C(d)}}\Bigr].
    \label{eq:omega_ge-approximation}
\end{equation}
Here, $\zeta(3)\approx1.2$ is the Riemann zeta function.

We first record the qubit frequency without the membrane as $\omega_q^\text{BFC}/2\pi = 4.93\ \text{MHz}$.
Inverting \refeq{eq:omega_ge-approximation} for known $E_J$ and $E_L$ yields $E_C(\infty)$.
Next, we simulate the qubit–membrane capacitance $C_m(d)$ versus gap $d$ using Ansys Q3D (see \reffig[(b)]{fig:distance-estimation})
Applying the definition from \refeq{eq:H_coupling}, we compute
\[
    E_C(d) = \frac{e^2}{\tfrac{e^2}{E_C(\infty)} + C_m(d)}.
\]
Substituting $E_C(d)$ back into \refeq{eq:omega_ge-approximation} gives $\omega_q(d)$ for fixed $E_J$, $E_L$, and $\omega_q^\text{BFC}$.
While $\omega_q^\text{BFC}$ is measured with high precision, we assign a 10 \% uncertainty to $E_J$ and $E_L$ and take their values found from the qubit spectroscopy in Sec.~\ref{sec:qubit-spectroscopy} of Methods.
Assigning 10 \% uncertainties to \(E_J\) and \(E_L\) and propagating them via a Monte Carlo analysis produces the shaded confidence band in \reffig[(a)]{fig:distance-estimation}.
After flip‐chip assembly, the qubit frequency drops to $\omega_q^\text{AFC}/2\pi = 2.35\ \text{MHz}$, from which we extract the gap distance $d = 2.5(3)~\upmu\text{m}$.

\section{Qubit Control and characterization}
\subsection{Qubit spectroscopy}
\label{sec:qubit-spectroscopy}
\begin{figure}
    \includegraphics{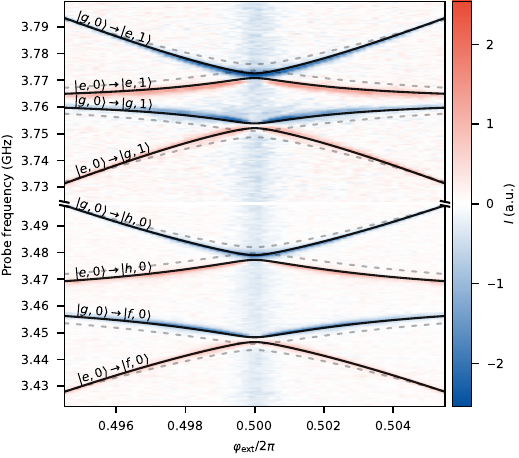}
    \caption{
        \textbf{Two‐tone spectroscopy of the heavy‐fluxonium qubit}: reflected readout amplitude versus probe frequency and external flux \(\varphi_\mathrm{ext}\). The primary “diamond” at \(\sim3.46\) GHz corresponds to the \(\{\ket{g},\ket{e}\}\to\{\ket{f},\ket{h}\}\) transitions. A secondary diamond near \(\sim3.65\) GHz, labeled by the transitions \(\{\ket{g,0},\ket{e,0}\}\to\{\ket{g,1},\ket{e,1}\}\) (chain‐mode Fock number in the ket), reveals coupling to a single chain mode. Solid lines are the best fit including that mode with parameters \(E_J/h =\text{4.886 GHz}\), \(E_C/h = \text{0.408 GHz}\), \(E_L/h = \text{0.135 GHz}\), \(\omega_{\rm chain}/2\pi = \text{3.650 GHz}\), and \(g_{qc}/2\pi = \text{197 MHz}\). Dashed lines show an alternative fit fixing \(\omega_{ge}/2\pi = 2.32\) MHz, yielding \(E_J/h = \text{4.757 GHz}\), \(E_C/h = \text{0.427 GHz}\), \(E_L/h = \text{0.121 GHz}\), \(\omega_{\rm chain}/2\pi = \text{3.641 GHz}\), and \(g_{qc}/2\pi = \text{203 MHz}\).
    }
    \label{fig:qubit-spectroscopy}
\end{figure}

The heavy‐fluxonium circuit, characterized by $E_J \gg E_C > E_L$, exhibits pronounced anharmonicity: its fundamental transition occurs at $\omega_q/2\pi = 2.35\ \mathrm{MHz}$, while next excited transitions lie in the GHz range.
To probe these higher transitions, we perform two‐tone spectroscopy: a variable‐frequency pump tone is applied to the fluxonium, followed by a readout pulse that captures the cavity response.
In conventional fluxonium devices, transitions between the lower manifold $(\ket{g}, \ket{e})$ and the upper manifold $(\ket{f}, \ket{h})$ produce a characteristic “diamond” pattern composed of four lines in the flux‐dependent spectrum, with the $\ket{g}\to\ket{f}$ and $\ket{e}\to\ket{h}$ transitions forbidden by symmetry at $\phiext = \pi$.
This pattern appears around 3.46~GHz (see \reffig{fig:qubit-spectroscopy}).

Our device exhibits, in addition, a second diamond-like structure around 3.7~GHz that the standard fluxonium Hamiltonian does not predict. We attribute this extra structure to coupling between the fluxonium and a collective mode of the junction chain employed as a superinductance~\citeS{Masluk2012}.
This bosonic mode, at $\omega_{\mathrm{chain}}$ is strongly coupled to the fluxonium, leading to a rich flux-dependent spectrum near $\varphi_\mathrm{ext} = \pi$.
The Hamiltonian of the coupled system, written in terms of the fluxonium flux and charge operators ($\op{\varphi}, \op{n}$) and the chain mode ladder operators ($\op{b}, \op{b}^\dagger$), is~\citeS{Viola2015}
\begin{equation}
    \begin{aligned}
        \op{H}
         & = 4E_C\,\op{n}^2 - E_J \cos\bigl(\op{\varphi} - \phiext\bigr) + \tfrac{1}{2}E_L\,\op{\varphi}^2                \\[6pt]
         & \quad + \hbar\omega_{\mathrm{chain}}\,\op{b}^\dagger \op{b} + \hbar g_{qc}\,(\op{b}^\dagger + \op{b})\,\op{n}.
    \end{aligned}
\end{equation}
We diagonalize this Hamiltonian numerically to obtain the eigenenergies and fit them to the experimental spectrum.

Beyond reproducing the extra diamond, the chain‐mode coupling also reshapes the original \(\{\ket{g},\ket{e}\}\to\{\ket{f},\ket{h}\}\) transitions of the fluxonium.
A fit to the transitions \(\{\ket{g},\ket{e}\}\to\{\ket{f},\ket{h}\}\) with the bare fluxonium model (not represented in the figure) yields a minimum gap of \(\omega_q/2\pi = 0.9 \ \mathrm{MHz}\) at \(\varphi_\mathrm{ext}=\pi\); including the chain-mode (full line fit of \reffig{fig:qubit-spectroscopy}) raises this to \(\omega_q/2\pi = 1.7\ \mathrm{MHz}\), in much closer agreement with the observed \(2.35\ \mathrm{MHz}\).
Although incorporating additional chain modes could further refine the theory-experiment agreement, we have chosen to keep a single chain-mode in the fit to keep the model tractable.
We also provide a best fit, where the $\ket{g}\to\ket{e}$ transition frequency is constrained around $2.35~\mathrm{MHz}$ (dashed lines in \reffig{fig:qubit-spectroscopy}). The quoted qubit parameters and uncertainties in the main text, $E_J/h = 4.82(6)$ GHz, $E_C/h = 0.418(9)$ GHz, $E_L/h = 0.128(7)$ GHz are taken as the averages and half‐differences of the two fitting results.

\subsection{Qubit reset and preparation fidelity}
\label{sec:qubit-preparation}

\begin{figure}
    \includegraphics{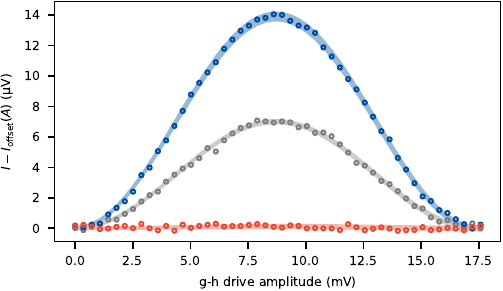}
    \caption[Qubit preparation fidelity.]{
        \textbf{Qubit preparation fidelity.} Rabi fringes obtained by driving the $\ket{g}\to\ket{h}$ transition after three different initialization sequences: $\ket{g}$ (blue), $\ket{e}$ (red), and thermal state (gray).
        The preparation fidelities \(\eta_g =99.5(1.7)\%\) and \(\eta_e=99.2(5)\%\) are extracted by comparing the contrast after \(\ket{g}\) and \(\ket{e}\) preparation to that obtained for a thermal state (solid lines; see ref.~\citeS{Najera2024} for details).}
    \label{fig:qubit-preparation}
\end{figure}
Because the \(\lvert g\rangle\leftrightarrow\lvert e\rangle\) transition (\(\omega_q/2\pi\approx2.35\)~MHz) is far below \(k_B T/h\approx0.2\)~GHz at 10~mK, the qubit thermalizes to an approximately 50:50 mixture of \(\lvert g\rangle\) and \(\lvert e\rangle\). In our previous work~\citeS{Najera2024}, sideband cooling via the readout resonator provided high‐fidelity initialization. In the present device, however, the narrower cavity linewidth (\(\kappa_r/2\pi = 240\)~kHz) prevents a sufficient cooling rate.

Instead, we employ optical pumping via a short‐lived collective mode of the junction chain at 6.42~GHz: to prepare \(\lvert e\rangle\), a continuous two‐photon drive at 3.2118~GHz transfers population from \(\lvert g,0\rangle\) to \(\lvert e,1\rangle\) (the second index is the chain‐mode photon number); rapid decay of that chain‐mode photon then leaves the qubit in \(\lvert e\rangle\). To prepare \(\lvert g\rangle\), we first pump into \(\lvert e\rangle\) as above and then apply a resonant \(\pi\) pulse on \(\lvert e\rangle\to\lvert g\rangle\). The \(\lvert e\rangle\)-preparation sequence is padded with a compensating delay after pumping so that both routines occupy the same total duration.

Preparation fidelity is evaluated using the diagnostic described in ref.~\citeS{Najera2024} Immediately after initialization, a single Rabi sweep on \(\lvert g\rangle\to\lvert h\rangle\) (152~ns pulse, varying amplitude) maps any \(\lvert g\rangle\) population into the auxiliary state \(\lvert h\rangle\). A 2~µs dispersive readout then distinguishes \(\lvert h\rangle\) from \(\{\lvert g\rangle,\lvert e\rangle\}\). Comparing the Rabi‐oscillation contrasts after three initializations—thermal equilibrium, \(\lvert g\rangle\)-preparation, and \(\lvert e\rangle\)-preparation (see \reffig{fig:qubit-preparation})—yields ground‐ and excited‐state fidelities \(\eta_g =99.5(1.7)\%\) and \(\eta_e=99.2(5)\%\) (see ref.~\citeS{Najera2024} for details).

\subsection{Single‐shot qubit readout}
\label{sec:qubit-readout}
\begin{figure}
    \includegraphics{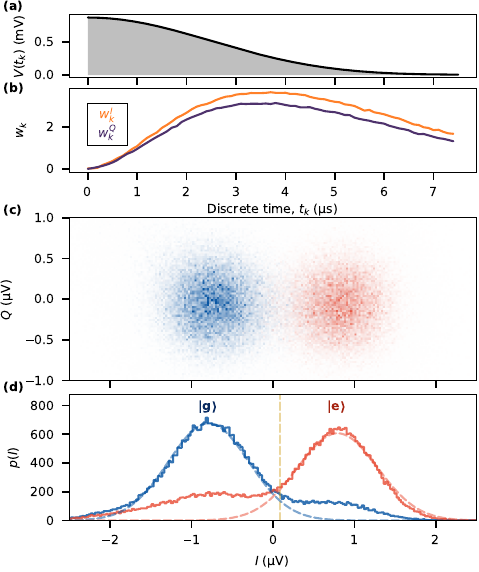}
    \caption{
        \textbf{Qubit readout fidelity.}
        \textbf{(a)} Amplitude envelope of the 7.5 µs readout pulse.
        \textbf{(b)} Time‐domain integration weights \(w_k^I\) and \(w_k^Q\) used to compute the \(I\)–\(Q\) quadratures: these weights are chosen as the optimal linear discriminant (matched filter) between the average \(\ket{g}\) and \(\ket{e}\) responses.
        \textbf{(c)} Histogram of single‐shot outcomes in the \(I\)–\(Q\) plane, collected after preparing \(\ket{g}\) (blue) or \(\ket{e}\) (red).
        \textbf{(d)} Histograms of the \(I\) quadrature for the same data. Colored dashed lines show the fitted Gaussian components (equal variance) that represent the ideal distributions in the absence of preparation and readout errors. The yellow vertical dashed line indicates the discrimination threshold chosen to maximize the contrast
        $P[I < T_I \mid \text{prep }e]  + P[I > T_I \mid \text{prep }g]$.}
    \label{fig:qubit-readout}
\end{figure}
We implement a direct, single‐shot discrimination of \(\lvert g\rangle\) versus \(\lvert e\rangle\) without transferring population to higher levels, in contrast to the approach previously used in ref.~\citeS{Najera2024}. The readout pulse is a 7.5~µs waveform with a smoothly decaying envelope (\reffig[(a)]{fig:qubit-readout}), designed to maximize separation of the \(\lvert g\rangle\) and \(\lvert e\rangle\) responses. During readout, the quadratures \((I_m, Q_m)\) are sampled every 4~ns, and final quadrature values \((I,Q)\) are computed via weighted integration using time‐dependent weights \(w^I_m,\,w^Q_m\) chosen as the optimal linear discriminant between the average \(\lvert g\rangle\) and \(\lvert e\rangle\) traces (\reffig[(b)]{fig:qubit-readout}).

To extract fidelities, we prepare the qubit in \(\ket{g}\) or \(\ket{e}\) and accumulate the single‐shot outcomes into \((I,Q)\) histograms (\reffig[(c)]{fig:qubit-readout}). A decision threshold \(T_I\) on \(I\) (dashed yellow line) is chosen to maximize Rabi‐oscillation contrast. The raw error probabilities are
\(
P[I < T_I \mid \text{prep }e] = 83.9(0.9)\%, \quad P[I > T_I \mid \text{prep }g] = 68.3(1.0)\%.
\)
Correcting for finite preparation fidelity (see ref.~\citeS{Najera2024}) yields true readout fidelities
\[
    P(I < T_I \mid e) = 84.8(1.8)\%, \quad P(I > T_I \mid g) = 69.1(1.2)\%,
\]
for an average fidelity of \(77.0(1.5)\%\). From the overlap of the two Gaussian peaks in the bi‐modal histogram (\reffig[(d)]{fig:qubit-readout}), we attribute a \(5.4\%\) error to misclassification. The remaining \(\sim17.6\%\) arises from thermal relaxation during the 7.5~µs readout pulse.

\subsection{Qubit AC Stark shift}
\label{sec:stark-shift}
\begin{figure}
    \includegraphics{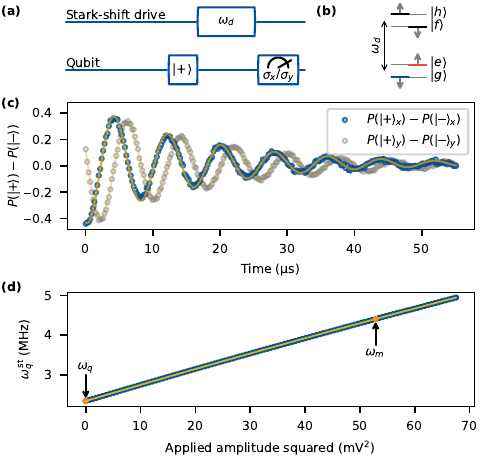}
    \caption{
        \textbf{Qubit AC–Stark shift.}
        \textbf{(a)} Ramsey pulse sequence with a continuous drive applied during the idle interval.
        \textbf{(b)} Energy‐level diagram of the four lowest fluxonium eigenstates.
        The drive frequency $\omega_d$ is set midway between the $\ket{e}\!\to\!\ket{f}$ and $\ket{g}\!\to\!\ket{h}$ transitions, inducing level repulsion and opposite AC–Stark shifts on the $\{\ket{g},\ket{h}\}$ and $\{\ket{e},\ket{f}\}$ manifolds.
        \textbf{(c)} Difference between the $\ket{+}$ and $\ket{-}$ measurement outcome probabilities along the $\sigma_x$ (blue) and $\sigma_y$ (gray) axes, measured at the bare
        qubit frequency.
        From these data, we reconstruct the complex transverse Bloch vector and fit it to exponentially decaying oscillations. The real and imaginary components of the fit are overlaid in yellow.
        \textbf{(d)} Qubit transition frequency extracted from \textbf{(c)} type of
        experiment as a function of the applied drive amplitude squared (proportional
        to $\Omega_d^2$). The yellow line shows a fit to  \refeq{eq:stark-shift-qubit-frequency}. The bare qubit frequency $\omega_q$ and membrane frequency $\omega_m$ are indicated by orange circles.
    }
    \label{fig:qubit-stark-shift}
\end{figure}

The lowest \(|g\rangle\!\leftrightarrow\!|e\rangle\) transition of our fluxonium lies at \(\omega_q/2\pi = 2.35\text{ MHz}\), while the mechanical mode sits at \(\omega_m/2\pi = 4.4\text{ MHz}\).
By applying a continuous drive at \(\omega_d/2\pi = 3.463\text{ GHz}\)—exactly midway between the \(\omega_\text{ef}\) and \(\omega_\text{gh}\) lines—we induce an AC–Stark shift that pulls the dressed qubit into resonance with the mechanics, at the cost of an increased dephasing rate (see~\refsec{sec:qubit-decoherence}).
To describe this driven system we therefore add a time‐dependent charge‐drive term to the fluxonium Hamiltonian:
\begin{equation}
    H = H_{\text{fluxonium}} + \hbar \Omega_d \cos(\omega_d t) \op{n},
    \label{eq:stark-shift-hamiltonian}
\end{equation}
where $\op{H_\text{fluxonium}}$ is the undriven fluxonium Hamiltonian and $\Omega_d$ is the coupling of the charge drive to the qubit.
We restrict our attention to transitions near $\omega_d$ and retain only the lowest four fluxonium levels, $\{\ket{g}, \ket{h}, \ket{e}, \ket{f}\}$.
Projecting the driving term of the Hamiltoninan into this four-level subspace, and considering only non-zero matrix elements at $\phiext = \pi$ yields
\begin{equation}
    \op{H}_{\mathrm{Stark}} = \hbar \Omega_d \cos{\omega_d t}\begin{pmatrix}
        0             & n_\text{gh} & n_\text{ge}   & 0           \\
        n_\text{gh}^* & 0           & 0             & 0           \\
        n_\text{ge}^* & 0           & 0             & n_\text{ef} \\
        0             & 0           & n_\text{ef}^* & 0
    \end{pmatrix}
\end{equation}
where $n_{ij}=\braket{i|\op{n}|j}$ is the charge matrix element.
We then transform into the frame rotating at $\omega_d$ via the unitary transformation
\begin{equation}
    \op{U} = \mathrm{diag}\!\begin{pmatrix}
        1 & e^{i\omega_{d}t} & e^{i \omega_\text{ge} t} & e^{i(\omega_{d} + \omega_\mathrm{ge}) t}
    \end{pmatrix}.
\end{equation}
In the rotated frame, dropping all fast rotating terms (beyond or close to $\omega_d$), the Hamiltonian writes
\begin{equation}
    \op{H}^\text{int} = \hbar \begin{pmatrix}
        0                        & \Omega_d n_\text{gh}/2 & 0                         & 0                       \\
        \Omega_d n_\text{gh}^*/2 & \Delta_\text{gh}       & 0                         & 0                       \\
        0                        & 0                      & 0                         & \Omega_d n_\text{ef} /2 \\
        0                        & 0                      & \Omega_d  n_\text{ef}^*/2 & \Delta_\mathrm{ef}
    \end{pmatrix},
\end{equation}
with $\Delta_\text{gh}= \omega_\text{gh} - \omega_d$ and  $\Delta_\text{ef}= \omega_\text{ef} - \omega_d$.
This Hamiltonian is block-diagonal, such that it can be diagonalized separately in the two subspaces $\{\ket{g}, \ket{h}\}$ and $\{\ket{e}, \ket{f}\}$.
The two lowest energy dressed eigenstates are
\begin{equation}
    \begin{aligned}
        \ket{\tilde{g}} & = \cos{\frac{\Theta_\text{gh}}{2}}\ket{g} - \sin{\frac{\Theta_\text{gh}}{2}}\ket{h}\,, \\
        \ket{\tilde{e}} & = \cos{\frac{\Theta_\text{ef}}{2}}\ket{e} - \sin{\frac{\Theta_\text{ef}}{2}}\ket{f}\,,
    \end{aligned}
    \label{eq:stark-shift-dressed-states}
\end{equation}
with mixing angles defined by $\tan\Theta_\text{gh} = \Omega_{d}n_\text{gh}/|\Delta_\text{gh}|$ and $\tan\Theta_\text{ef} = \Omega_{d}n_\text{ef}/|\Delta_\text{ef}|$.
The eigenerergies are given by
\begin{equation}
    \begin{aligned}
        \tilde{E}_{g}/\hbar
         & = \frac{1}{2}\,\text{sign}\bigl(\Delta_{\rm gh}\bigr)\,\bigl(|\Delta_{\rm gh}| - \sqrt{\Delta_{\rm gh}^2+\Omega_d^2\,n_{\rm gh}^2}\bigr)\,, \\
        \tilde{E}_{e}/\hbar
         & = \frac{1}{2}\,\text{sign}\bigl(\Delta_{\rm ef}\bigr)\,\bigl(|\Delta_{\rm ef}|-\sqrt{\Delta_{\rm ef}^2+\Omega_d^2\,n_{\rm ef}^2}\bigr)\,.
    \end{aligned}
\end{equation}
For the circuit at half‐flux, $\phiext = \pi$, one finds $n_\text{gh}\simeq n_\text{ef}\equiv n_\text{up}$, furthermore, we have chosen a drive frequency precisely midway between $\omega_\text{gh}$ and $\omega_\text{ef}$ yielding $\Delta_\text{gh} = -\Delta_\text{ef} \equiv \Delta_\text{st}$, and $\Theta_\text{gh}\simeq\Theta_\text{ef}\equiv\Theta$.
In this approximation, the AC Stark shift is given by
\begin{equation}
    \begin{aligned}
        \delta\omega_{q}
         & \equiv \tilde{E}_{e} - \tilde{E}_{g}                                     \\
         & =\sqrt{\Delta_\text{st}^2 + \Omega_d^2 n_\text{up}^2} - \Delta_\text{st} \\
         & \approx \frac{\Omega_d^2 n_\text{up}^2}{2\Delta_\mathrm{st}}
        \left(1- \frac{1}{4}\frac{\Omega_d^2 n_\text{up}^2}{\Delta_\mathrm{st}^2}\right).
    \end{aligned}
    \label{eq:stark-shift-qubit-frequency}
\end{equation}
In the last line, we expanded the expression to fourth order in $\Omega_d/\Delta_\mathrm{st}$.

Since the absolute attenuation between source and sample is unknown, we calibrate the AC–Stark shift by Ramsey interferometry (\reffig[(a)]{fig:qubit-stark-shift}).
Beginning at the fluxonium sweet-spot frequency \(\omega_q/2\pi\approx2.35\) MHz, we apply a continuous tone at \(\omega_d/2\pi=3.463\) GHz with varying source power \(P\).
A \(\pi/2\)–Ramsey sequence maps the drive-induced phase into \(\sigma_z\), from which we extract the dressed qubit frequency.
Plotting this frequency versus \(P\) reveals a primarily linear Stark pull plus a small quadratic correction (\reffig[(c)]{fig:qubit-stark-shift}).
Because \refeq{eq:stark-shift-qubit-frequency} fixes the ratio of the linear and quadratic terms, only one free parameter—the scaling factor between \(P\) and the on-chip Rabi rate \(\Omega_d\)—is required to fit all the data.
Over the accessible power range the qubit frequency tunes continuously from 2.35 MHz up to $\sim$ 5 MHz (more than one octave), enabling resonance with the 4.4 MHz mechanical mode.
The resulting single-parameter fit (solid line, \reffig[(c)]{fig:qubit-stark-shift}) quantitatively reproduces both slope and curvature of the Stark shift.

\subsection{Decoherence and relaxation times}
\label{sec:qubit-decoherence}

\begin{figure}
    \includegraphics{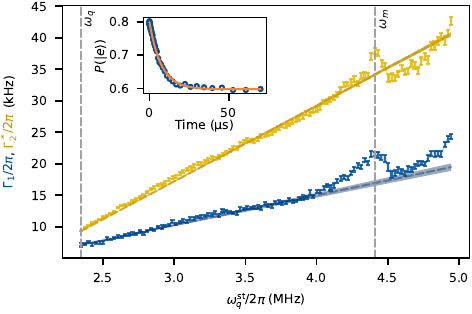}
    \caption{
        \textbf{Qubit decoherence and relaxation rates.}
        Energy‐relaxation rate \(\Gamma_1 = 1/T_1\) (blue markers) and free‐induction‐decay rate \(\Gamma_2^* = 1/T_2^*\) (yellow markers) are plotted versus the AC‐Stark–shifted qubit frequency \(\omega_q^\mathrm{st}\).
        \(\Gamma_1\) is extracted from the exponential relaxation of the qubit population toward its thermal equilibrium.
        \(\Gamma_2^*\) is obtained from Ramsey‐fringe decay.
        The dashed blue line shows \(\Gamma_1(\omega_q^\mathrm{st})\) from \refeq{eq:stark-shift-gamma-1}, with \(\Gamma_1^{ge}\) as the sole fit parameter.
        The dashed yellow line represents \(\Gamma_2^* = \Gamma_1/2 + \Gamma_\phi\), using \(\Gamma_\phi(\omega_q^\mathrm{st})\) from \refeq{eq:stark-shift-gamma-phi} and fitting \(\delta\Omega_d/\Omega_d\).
        Vertical black dashed lines indicate the bare qubit frequency $\omega_q$ and the mechanical mode frequency $\omega_m$.
        Inset: Qubit relaxation at $\omega_m$ following preparation in the state $\ket{e}$.
        Experimental data (blue) and corresponding exponential fit (orange) are shown.
        \label{fig:qubit-t1-t2}
    }
\end{figure}

We measure the qubit relaxation and dephasing rates as functions of the Stark‐shifted qubit frequency, $\omega_q^\mathrm{st}$.
The energy‐relaxation rate, $\Gamma_1 = 1/T_1$, is extracted from the exponential decay of the excited‐state population.
The free‐induction‐decay rate, $\Gamma_2^* = 1/T_2^*$, is obtained from a Ramsey experiment (\reffig[(a)]{fig:qubit-stark-shift}).
The resulting rates are plotted in \reffig{fig:qubit-t1-t2}.
At the bare qubit frequency we observe $\Gamma_1/\Gamma_2^* \simeq 1.12<2$.
As the AC–Stark drive increases, the pure dephasing rate $\Gamma_{\phi} = \Gamma_2^* - \Gamma_1/2$ rises sharply, causing $T_2^*$ to degrade much faster than $T_1$.
Both rates exhibit a pronounced peak when $\omega_q^\mathrm{st}$ approaches $\omega_m$, due to the qubit transitions induced by the thermally excited mechanical mode.

We attribute the approximately linear rise of \(\Gamma_1\) with detuned qubit frequency \(\omega_q^\mathrm{st}\) to the growing admixture of the short‐lived \(\{\ket{f},\ket{h}\}\) states into the lower dressed states—an effect analogous to dressed‐state spontaneous emission in optical tweezers (see~\refeq{eq:stark-shift-dressed-states}).  If the bare‐manifold decay rates are \(\Gamma_1^{ge}\) and \(\Gamma_1^{fh}\), then the dressed‐state relaxation rate is
\[
    \Gamma_1(\delta\omega_q)
    =\cos^2\!\bigl(\tfrac{\Theta}{2}\bigr)\,\Gamma_1^{ge}
    +\sin^2\!\bigl(\tfrac{\Theta}{2}\bigr)\,\Gamma_1^{fh}\,.
\]
For small mixing angle \(\Theta \approx \delta\omega_q/\Delta_{\rm st}\), expanding to first order gives
\begin{equation}
    \label{eq:stark-shift-gamma-1}
    \Gamma_1(\delta\omega_q)
    \approx \Gamma_1^{ge}
    +\frac{\Gamma_1^{fh}}{2\,\Delta_{\rm st}}\,\delta\omega_q\,.
\end{equation}
The dashed blue line in \reffig{fig:qubit-t1-t2} plots \refeq{eq:stark-shift-gamma-phi} using the independently measured value \(\Gamma_1^{fh}/2\pi = 90\)~kHz, leaving only \(\Gamma_1^{ge}\) as a fit parameter.

We attribute the enhanced dephasing rate \(\Gamma_2^*(\omega_q^\mathrm{st})\) to amplitude fluctuations of the Stark drive \(\Omega_d\), yielding
\begin{equation}
    \label{eq:stark-shift-gamma-phi}
    \Gamma_{\phi}
    = \frac{d(\delta\omega_q)}{d\Omega_d}\,\delta\Omega_d
    \simeq 2\,\delta\omega_q\,\frac{\delta\Omega_d}{\Omega_d}\,,
\end{equation}
where the approximation holds for \(\Omega_d\ll\Delta_{\rm st}\). Assuming classical amplitude noise (i.e.\ constant \(\delta\Omega_d/\Omega_d\)), fitting this expression to the data (yellow dashed line in \reffig{fig:qubit-t1-t2}) gives \(\delta\Omega_d/\Omega_d\approx0.6\%\), in good qualitative agreement with the \(\sim0.2\%\) specification of our Anapico source.
 \section{Spectrum analyzer theory}
\label{sec:spectrum_analysis}
\label{sec:cooling_heating}

\subsection{Quantum harmonic oscillator spectrum}\label{sec:spec_theory}

The mechanical system is modeled as a harmonic oscillator, with annihilation operator \(\aop\), coupled to a thermal bath at rate \(\kappa_m\) yielding a mean thermal phonon number \(\nmth\).
The Hamiltonian and loss operators capturing the system dynamics can be written as
\begin{gather*}
    \op{H}_m/\hbar = \left(\omega_m + \frac{1}{2}\right) \aopd \aop, \\
    \op{L}_{m,\downarrow} = \sqrt{\kappa_m (n_{th}+1)}\,\aop,\quad
    \op{L}_{m,\uparrow}   = \sqrt{\kappa_m\,n_{th}}\,\aopd.
\end{gather*}
The evolution of the system is modeled by the Lindblad equation \(\dot{\rhop} = \sop{L}_m\rhop\), with
\begin{equation}
    \sop{L}_m\cdot = -i[\op{H}, \cdot] + \sop{D}[\op{L}_{m,\downarrow}]\cdot + \sop{D}[\op{L}_{m,\uparrow}]\cdot
\end{equation}
the Lindbladian.

Closely following the reasoning of ref.~\citeS{Clerk2010}, we'll discuss in this part the quantum noise spectrum.
Similarly to classical physics, the noise spectrum of the Heisenberg picture position operator \(\op{x}(t) = \op{x}e^{\sop{L}_mt}\) can be written as
\begin{equation}
    \mathcal{S}_{\op{x}\op{x}}(\omega) = \int_{-\infty}^{+\infty}dt \; e^{i\omega t}\braket{\op{x}(t)\op{x}(0)}.
\end{equation}
In general, for \(t\neq 0\), the commutator \([\op{x}(t), \op{x}(0)]\) is a complex non-zero quantity.
This reflects the fact that the measurement of \(\op{x}(t)\) is necessarily impacted by the back-action of the measurement of \(\op{x}(0)\).
Consequently, the complex-valued auto-correlator \(\mathcal{C}_{\op{x}\op{x}}(t) = \braket{\op{x}(t)\op{x}(0)}\) is a purely mathematical construction that cannot be evaluated by a simple continuous measurement of $\op{x}(t)$.
Indeed, in a practical measurement setting, the added quantum noises ensure that the measured quantities $\op{x}(t)$ commute with each other at all times $t$, resulting in real-valued correlators.
Nevertheless, the complex-nature of \(\mathcal{C}_{\op{x}\op{x}}(t)\), which results in a frequency asymmetric spectrum \(S_{\op{x}\op{x}}(\omega)\), has a profound physical meaning.
We can gain insights by developing \(\op{x}(t) = (\aopd + \aop)/2\), in it's elementary components and dropping vanishing terms
\begin{equation}
    \braket{\op{x}(t)\op{x}(0)} = \frac{1}{4}\Bigl(\braket{\aop(t)\aopd(0)} + \braket{\aopd(t)\aop(0)}\Bigr).
\end{equation}
The first correlator \(\braket{\aop(t)\aopd(0)}\) can be understood as the amplitude of probability in a fictitious experiment where an excitation would be recovered at time \(t\) after being added to the oscillator at time \(0\).
Therefore, the noise spectrum associated to this correlator is the absorption spectrum of the mechanical mode.
Similarly, the noise spectrum associated to \(\braket{\aopd(t)\aop(0)}\) is the emission spectrum of the mechanical mode, where an excitation is removed from the oscillator at time \(0\) and re-injected at time \(t\).
Since the mechanical mode is coupled to a bath, both correlators tend to $0$ exponentially as the excitation can be lost in the interval.
Computing this correlators explicitly, we obtain
\begin{equation}
    \begin{aligned}
        \braket{\op{x}(t)\op{x}(0)}
        = \frac{1}{4}\Bigl(
         & \braket{\aopd(0)\aop(0)} e^{-i\omega_m t} + \\
         & + \braket{\aop(0)\aopd(0)}e^{i\omega_m t}
        \Bigr)e^{-\frac{\kappa_m}{2}|t|}.
    \end{aligned}
\end{equation}
Which in turn yields the spectra
\begin{gather}
    \mathcal{S}_{\op{x}\op{x}}(\omega) = \mathcal{S}_{\aopd\aop}(\omega) + \mathcal{S}_{\aop\aopd}(\omega)
\end{gather}
with
\begin{equation}
    \begin{aligned}
        S_{\aopd\aop}(\omega) & = \frac{\nmth\kappa_m}{(\omega-\omega_m)^2 + (\kappa_m/2)^2},     \\
        S_{\aop\aopd}(\omega) & = \frac{(\nmth+1)\kappa_m}{(\omega+\omega_m)^2 + (\kappa_m/2)^2},
    \end{aligned}
    \label{eq:ap_emission_absorption_spectra}
\end{equation}
where the negative and positive frequency parts corresponds to the emission and absorption spectrum, respectively.
Crucially, the total spectrum \(\mathcal{S}_{\op{x}\op{x}}(\omega)\) is not symmetric-in-frequency and cannot be obtained from the correlator of a real-valued time-dependent observable \(\braket{\op{A}(t)\op{A}(0)}\).
Instead, we separately measure the real part of the autocorrelation functions \(\mathrm{Re}(\mathcal{C}_{\aopd\aop})\) and \(\mathrm{Re}(\mathcal{C}_{\aop\aopd})\), the Fourier transform of which yields the symmetrized spectra $\meanspec{\aopd \aop}$ and $\meanspec{\aop \aopd}$, by inspecting the correlation between successive measurements where the qubit is repeatedly prepared in \(\ket{g}\) or \(\ket{e}\) respectively.

\subsection{System definition}

In this section we explain how to experimentally measure the emission and absorption spectrum from periodic weak measurements of the membrane with a qubit.
The system is composed of a Fluxonium, modeled as a qubit with frequency \(\omega_q\) and the membrane mode.
The two systems interact with a Jaynes-Cummings Hamiltonian \(\op{H_c}/\hbar = i \tfrac{\Omega}{2} (\aop + \aopd)(\op{\sigma} - \op{\sigma}^\dag)\), yielding the system Hamiltonian
\begin{equation*}
    \op{H} = \frac{\hbar\omega_q}{2}\op{\sigma_z} + \op{H}_m + \op{H}_c.
\end{equation*}

In the spectrum analyzer experiment, the qubit frequency is regularly Stark-shifted, close to the membrane frequency, to \(\omega_m - \Delta \).
In the frame rotating at the Stark-shifted qubit frequency for both the membrane and the qubit, after a rotating-wave approximation, the Hamiltonian simplifies to \(\op{H}/\hbar = \Delta \aopd \aop + i \tfrac{\Omega}{2} (\aopd\op{\sigma} - \aop\op{\sigma}^\dag)\).
When the qubit frequency is not shifted, we can neglect the effect of the coupling.
The Lindbladian superoperator of the membrane free evolution is \(\sop{L}_m\rhop = -i[\Delta \aopd \aop, \rhop] + \sop{D}[\op{L_{m,\downarrow}}]\rhop + \sop{D}[\op{L_{m,\downarrow}}]\rhop\).
For the sake of simplicity, we will consider for now an ideal qubit and we will treat the qubit imperfections in \refapp{app:sideband_asymmetry_imperfect}.

\subsection{Semi-classical treatment}
\label{app:semiclassical}
\tocless{\subsubsection}{Autocorrelation function}
In the semi-classical hypothesis, we treat classically the mechanical mode degrees of freedom, \(\op{x}\) and \(\op{p}\).
In the frame of the qubit, the mode complex amplitude \(\alpha(t)\) is slowly oscillating at rate \(\Delta\) with random phase and amplitude fluctuations over a characteristic time-scale \(1/\kappa_m\).
As the interaction time is short compared to the mechanical mode time-scales, \(\alpha(t)\) can be considered constant over one interaction.
Moreover, the interaction Hamiltonian can be simplified to \(\op{H}_c/\hbar = i\tfrac{\Omega}{2}( \alpha^*\op{\sigma} - \alpha\op{\sigma}^\dag)\), which is a classical drive with amplitude \(\alpha\) on the qubit.
Expanding the terms, we obtain that \(\op{H}_c/\hbar = \tfrac{\Omega}{2} \left(\mathrm{Im}(\alpha)\op{\sigma}_x + \mathrm{Re}(\alpha)\op{\sigma}_y\right)\).
The unitary evolution after an interaction time \(\tau\) can be written as
\begin{equation}
    \op{U}_I
    = \exp\Bigl(-\frac{i}{\hbar} \op{H}_c\tau\Bigr)
    = \exp\Bigl(-i\frac{\Omega}{2} |\alpha|\vec{n}\cdot\vec{\op{\sigma}}\tau\Bigr) \\
\end{equation}
where
\(\vec{n} = \frac{1}{|\alpha|}(\mathrm{Im}(\alpha),\mathrm{Re}(\alpha),0)\) and
\(\vec{\op{\sigma}}=(\op{\sigma}_x,\op{\sigma}_y,\op{\sigma}_z)\).
Comparing with the standard \(SU(2)\) rotation matrix \(\exp(-i\tfrac{\theta}{2} \vec{n} \cdot \vec{\op{\sigma}})\) shows that the qubit undergoes a rotation by \(\theta = \Omega |\alpha|\tint\) about the axis \(\vec{n}\) in the Bloch‐sphere equatorial plane, whose complex direction is \(i\alpha^*\).
Therefore, measuring \(\op{\sigma}_x\) of the qubit, at time \(t_k\), performs a measurement of the position of the mechanical mode as
\begin{equation}
    \braket{\op{\sigma}_x(t_k)} = \theta_1 x(t_k).
    \label{eq:expectation_semi_classical}
\end{equation}
In this semi-classical treatment, where the position $x(t)$ is an external parameter which is not subject to qubit backaction, the measurement outcomes at different times are independent quantum observables. We thus have $\braket{\op{\sigma}_{x,k} \op{\sigma}_{x,{k'}}} = \braket{\op{\sigma}_{x, k}} \braket{\op{\sigma}_{x, {k'}}}$ for $k\neq k'$. It thus follows from \refeq{eq:expectation_semi_classical} that  \begin{equation}
    \label{eq:semi_cla_auto}
    \mathcal{C}_k \equiv \overline{\braket{\op{\sigma}_{x,k} \op{\sigma}_{x,{0}}}} = \theta_1^2\overline{x(t_k)x(0)},
\end{equation}
where the overline denote the averaging over the different trajectories of the harmonic oscillator.
Moreover, in the case where $k=0$, the autocorrelation reads \(\mathcal{C}_0 = 1\) as \(\sigma_x^2 = \II\).
This proves \refeq{eq:autocorrelator-factorization} of the main text.

\tocless{\subsubsection}{Wiener-Khintchine theorem}

Let us define the discrete Fourier transform of a signal \(\{m_k\}_{0\leq k< N}\) sampled at a frequency \(1/T\) as \(\mathcal{F}[\{m_k\}](\omega_r) = \sum_{n=0}^{N-1}m_n e^{-i\omega_r n}\), where \(\omega_r = 2\pi r/NT\) for \(0\leq r<N\).
These discrete time frequencies maps unambiguously to continuous time frequencies \(\omega\) on the first Nyquist zone \(0\leq\omega<\pi/T\). We thus drop the subscript $r$ in the following.
We define the power spectral density as
\begin{equation}
    S_{mm}(\omega) = \tcycle\lim_{N\to\infty}\frac{1}{N}\overline{|\mathcal{F}[\{m_k\}](\omega)|^2}.
\end{equation}
The averaged squared modulus of the Fourier transforms can be expressed as
\begin{equation}
    \label{eq:averaged_fft}
    \begin{aligned}
        \overline{|\mathcal{F}[\{m_k\}](\omega)|^2} = & \overline{\Bigl(\mathcal{F}[\{m_k\}](\omega)\Bigr)^*\mathcal{F}[\{m_k\}](\omega)} \\
        =                                             & \sum_{k = 0}^{N-1}\sum_{l = 0}^{N-1} \overline{m_k m_l} e^{-i\omega (k-l)}
    \end{aligned}
\end{equation}
By stationarity of the measurement signal \(\overline{m_k m_l} = \overline{m_{k-l} m_0} \equiv \mathcal{C}_{k-l}\), defining \(\mathcal{C}_{r}\) as the autocorrelation function at time \(t_r = rT\).
Changing in \refeq{eq:averaged_fft} the sum index \(k\) to \(r=k-l\), we obtain
\begin{equation*}
    \overline{|\mathcal{F}[\{m_k\}](\omega)|^2}
    = \sum_{r = -N+1}^{N-1}\sum_{l = f(r)}^{g(r)} \mathcal{C}_r e^{-i\omega r},
\end{equation*}
where \(f(r)=\mathrm{max}(0, r)\) and \(g(r) = f(r) + N-1-|r|\).
The second sum is composed of \(N-|r|\) terms independent of \(l\).
Taking the limit of this sum leads to the result
\begin{equation}
    \begin{aligned}
        \frac{1}{N}\overline{|\mathcal{F}[\{m_k\}](\omega)|^2}
         & = \sum_{r = -N+1}^{N-1} \mathcal{C}_r e^{-i\omega r} \left(1-\frac{|r|}{N}\right) \\
         & \xrightarrow[N \to \infty]{} \sum_{r = -N+1}^{N-1} \mathcal{C}_r e^{-i\omega r}.
    \end{aligned}
\end{equation}
The inversion of the sum and limit in the last line is valid in the hypothesis of an absolute summable autocorrelation, i.e. a power spectral density that does not diverge.
This proves the Wiener-Khintchine theorem, linking the power spectral density to the Fourier transform of the autocorrelation function.
\begin{equation}
    S_{mm}(\omega) = \tcycle\mathcal{F}[\{\mathcal{C}_r\}](\omega).
\end{equation}
Using \refeq{eq:semi_cla_auto}, this theorem is used to prove \refeq{eq:spectra_semi_classical} of the main text.

\subsection{Quantum treatment}

\tocless{\subsubsection}{Measurement operators}

The qubit being initialized in the pure state $\ket{g}$ or $\ket{e}$ before each interaction, we can safely assume that the initial joint qubit-membrane state is described by the factorized density operator \(\op{\rho_m} \otimes \ket{i}\bra{i}\) (with $i\in\{g, e\}$).
In the rotating frame, neglecting the small qubit-membrane detuning $\Delta$ and decoherence during the short interaction time $\tau$, the evolution operator $\op{U}_I=e^{-i \theta_1/2 (\aop^\dagger \op\sigma - \aop \op\sigma^\dagger)}$ writes~\citeS{Rouchon2022}
\begin{equation}
    \op{U_I} =  \begin{pmatrix}
        \cos\left(\frac{\theta_1}{2}\sqrt{\aopd\aop+1}\right)                                 &
        -\aop\frac{\sin\left(\frac{\theta_1}{2}\sqrt{\aopd\aop}\right)}{\sqrt{\aopd\aop}}                                                           \\
        \aopd\frac{\sin\left(\frac{\theta_1}{2}\sqrt{\aopd\aop+1}\right)}{\sqrt{\aopd\aop+1}} & \cos\left(\frac{\theta_1}{2}\sqrt{\aopd\aop}\right)
    \end{pmatrix},
\end{equation}
where the matrix is written in the qubit basis \(\{\ket{e},\ket{g}\}\). During the interaction, the system state evolves according to:
\begin{equation}
    \op{\rho}(t_k) \longrightarrow \op{\rho}(t_{k+1}) = \op{U}_I \left(\op{\rho}_m(t_k) \otimes \ket{i}\bra{i} \right)\op{U}_I^\dagger.
\end{equation}
In the final state $\op{\rho}(t_{k+1}) $, we perform a projective measurement of the qubit in the $\ket{\pm}$ basis. The probability to measure the qubit in state $\ket{+}$ or $\ket{-}$ is then
\(
p_{\pm,i} (t_{k+1}) = \text{Tr} [\op{\rho}(t_{k+1}) \ket{\pm}\bra{\pm}]= \text{Tr}[\op{M}_{\pm,i} \rhop_m(t_k) \op{M}_{\pm,i}]
\), where we defined the measurement operators
\begin{equation}
    \begin{aligned}
        \op{M}_{\pm, i}
         & = \bra{\pm}\op{U}_I\ket{i} \\
         & \simeq \frac{1}{\sqrt{2}}
        \begin{cases}
            \II \mp \frac{\theta_1}{2} \aop - \frac{\theta_1^2}{8} \aopd \aop  & \text{if } i=g, \\
            \II \pm \frac{\theta_1}{2} \aopd - \frac{\theta_1^2}{8} \aop \aopd & \text{if } i=e.
        \end{cases}
    \end{aligned}
\end{equation}
The final expression is calculated up to second order in $\theta_1$, and we removed an irrelevant global phase term in the \(i=e\) case.
The set of measurement operators describes a Positive Operator-Valued Measure which fulfills $\op{M}_{+, i}^\dag \op{M}_{+, i} + \op{M}_{-, i}^\dag \op{M}_{-, i} = \II$ up to second order in $\theta_1$.
Moreover, since \(\op{M}_{\pm, i}\) are close to the identity for \(\theta_1 \ll 1\), this process describes more specifically a weak measurement.
Upon qubit measurement, the membrane density matrix is projected to
\begin{equation}
    \label{eq:measurement_operators}
\op{\rho}_m^{\pm, i} (t_{k+1}) = \frac{\op{M}_{\pm, i}\rhom(t_k)\op{M}_{\pm, i}^\dag}{p_{\pm, i}} \equiv \frac{\Kr_{\pm, i}\rhom(t_k)}{p_{\pm, i}}.
\end{equation}
Here, we defined the measurement Kraus map $\Kr_{\pm, i}$.
Introducing the dissipation and measurement superoperators as
\[
    \sop{D}[\op{L}]\,\op{\rho} = \op{L}\,\op{\rho}\,\op{L}^\dagger - \tfrac12\{\op{L}^\dagger \op{L},\op{\rho}\},
    \qquad
    \sop{M}[\op{L}]\,\op{\rho} = \op{L}\,\op{\rho} + \op{\rho}\,\op{L}^\dagger,
\]
the Kraus map can be expressed, up to second order in $\theta_1$, as
\begin{equation}
    \label{eq:kraus}
    \sop{K}_{\pm,i}
    =\frac12\Bigl[\II \pm \frac{\theta_1}{2}\,\sop{M}[\op{L}_i]
        + \frac{\theta_1^2}{4}\,\sop{D}[\op{L}_i]\Bigr],
\end{equation}
where the weakly measured membrane operators, for each qubit state preparation are
\begin{equation}
    \op{L}_i = \begin{cases}
        -\aop, & i = g, \\
        \aopd, & i = e.
    \end{cases}
\end{equation}
The term proportional to \(\sop{M}[L]\) in \refeq{eq:kraus} is the outcome‐dependent “informational” (or quantum) backaction, which steers the mechanical state based on the measurement result, while the \(\sop{D}[L]\) contribution describes a measurement‐independent, qubit‐induced damping channel, directly analogous to dynamical backaction in cavity optomechanics~\citeS{Aspelmeyer2014}.

\tocless{\subsubsection}{Dynamical back-action}
\label{subsec:dynamical_backaction}

When the measurement outcomes are disregarded, the membrane evolves under the average Kraus map
\begin{equation}
    \label{eq:kraus_mean}
    \bar{\sop{K}}_i = \sop{K}_{+,i} + \sop{K}_{-,i}
    = \II + \frac{\theta_1^2}{4}\,\sop{D}[\op{L}_i].
\end{equation}
During the interval $T$ between two successive qubit measurements, the membrane undergoes free-evolution and dissipation to the thermal bath, described by the Lindbladian
\begin{equation}
    \sop{L}_{m} =
    -i[\Delta \aopd \aop,\cdot]
    + \kappa_m \nmth\,\sop{D}[\op{a}]
    + \kappa_m(\nmth+1)\,\sop{D}[\op{a}^\dag].
\end{equation}
Interleaving the average Kraus map \eqref{eq:kraus_mean} with the membrane’s free evolution $e^{\sop{L}_m \,T}$ yields
\begin{equation}
    \op{\rho}(t_k + T) =
    e^{\sop{L}_m \,T} \;\bar{\sop{K}}_i\,\op{\rho}(t_k)
    \approx
e^{\sop{L}_{m,i}\,T},
\end{equation}
where, in a coarse-grained approximation, the effective Lindbladian becomes
\(
\sop{L}_{m,i}
= \sop{L}_m + \kappa\,\sop{D}[\op{L}_i]
\), with the dynamical‐backaction rate \(\kappa=\theta_1^2/4T\).
This coarse-grained evolution is identical to that induced by the coupling at a rate $\kappa_{m, i}'$ to an effective environment with thermal occupation $n_\mathrm{th, i}'$, where
\[
    \kappa'_{m,i} =
    \begin{cases}
        \kappa_m + \kappa, & i=g, \\
        \kappa_m - \kappa, & i=e,
    \end{cases}
    \quad
    n'_{\text{th,i}} =
    \frac{\kappa_m n_{\rm th} + \delta_{e,i}\,\kappa}{\kappa'_{m,i}}.
\]
Here \(\delta_{e,i}=1\) if \(i=e\) (zero otherwise).  In other words, dynamical back‐action modifies both the damping rate and the apparent thermal occupation by factors \(1\pm C\), where \(C=\kappa/\kappa_m\) is the cooperativity, in line with conventions commonly used in optomechanics.
The last additive term in the occupation \(\kappa / \kappa_m' = C/(1-C)\) appearing when the qubit is prepared in state $\ket{e}$ corresponds to the additional fluctuations resulting from quantum backaction. In our experiment where the cooperativity is on the order of 10 \%, we can safely neglect this contribution in the area of the spectrum $S_{mm}^e(\omega)$.

Finally, if we alternate ground and excited preparations each cycle, the net evolution over a full cycle is generated by
\begin{equation}
    \sop{L}_{m, ge}
    = \sop{L}_m + \kappa\,\sop{D}[\aop] + \kappa\,\sop{D}[\aopd],
\end{equation}
so that
\[
    \kappa'_{m, ge} = \kappa_m,
    \qquad
    n'_{\text{th}, ge} = n_{\rm th} + C.
\]
Thus the alternating protocol cancels the first‐order dynamical back‐action, leaving only a negligible residual heating of order \(C\ll n_{\rm th}\).
Finally, note that here, the interval between two identical preparation steps \(T\) is effectively doubled (see \reffig[(b)]{fig:spec_asym}) compared to the previous protocol, which halves both the back‐action rate \(\kappa\) and the cooperativity \(C\).

\tocless{\subsubsection}{Emission and absorption spectra}
\label{app:emission_spectra}

The full Kraus operators factor into the mean‐evolution map and a pure backaction part:
\begin{equation}
    \Kr_{\pm, i} = \bar{\Kr}_i\Kr_{\pm, i}' \; ,
    \text{ with }
    \Kr_{\pm, i}' =
    \frac{1}{2}
    \left(\II \pm \frac{\theta_1}{2} \sop{M}[\op{L}_i] \right).
\end{equation}
One can then calculate the joint probability to obtain outcomes \(m_0\) and \(m_k\) separated by \(k\) full cycles becomes
\begin{align}
    \label{eq:conditional_probability_mk_m0}
    \mathbb{P}[m_k, m_0|\rhop_m, i] \simeq \mathrm{Tr}[\Kr_{m_k, i}' e^{\widetilde{\sop{L}}_{m} k\tcycle}\Kr_{m_0, i}' \rhop_m],
\end{align}
where $\sop{\widetilde{L}}_{m}$ denotes the effective Lindbladian for the chosen preparation sequence—either $\sop{L}_{m,g}$ or $\sop{L}_{m,e}$ for identical preparations, or $\sop{L}_{m, ge}$ for alternating ground/excited protocols.

We then express the discrete‐time autocorrelation of the measurement outcomes as
\[
    \mathcal{C}_k^i \;=\;\sum_{m_0,m_k=\pm1} m_k\,m_0\;
    \mathbb{P}[m_k,m_0\mid\rhop_m,i],
\]
so that for \(k=0\) trivially \(\mathcal{C}_0^i=1\).  For \(k\neq0\), inserting the expression \eqref{eq:conditional_probability_mk_m0}, and using
\(\sum_{m=\pm1}m\,\sop{K}'_{m,i}=\tfrac{\theta_1}{2}\,\sop{M}[L_i]\)
yields
\[
    \mathcal C_k^i
    = \frac{\theta_1^2}{4}\,
    \Tr\bigl[\sop{M}[\op{L}_i]\,
        e^{\tilde{\sop{L}}_m\,k T}\,
        \sop{M}[\op{L}_i]\,\rhop_m\bigr].
\]
Defining the Heisenberg picture operator $\op{O}(t)$ as
\(\op{O}\rhop(t) = \op{O}e^{\widetilde{\sop{L}}_m t}\rhop = \op{O}(t)\rhop \),
the measurement correlator reads
\[
    \mathcal C_k^i\big|_{k\neq0}
    = \frac{\theta_1^2}{4}\,
    \text{Re}\bigl\langle \op{L}_i^\dagger(t_k)\,\op{L}_i(0)\bigr\rangle,
\]
The discrete‐time spectrum then follows by Fourier transform.
\begin{equation}
    \begin{aligned}
        \mathcal{S}_{mm}^g(\omega)   & = \tcycle + \frac{\theta_1^2}{4}\bar{\mathcal{S}}_{\aopd\aop}(\omega), \\
        \mathcal{S}_{mm}^e(\omega_k) & = \tcycle + \frac{\theta_1^2}{4}\bar{\mathcal{S}}_{\aop\aopd}(\omega),
    \end{aligned}
\end{equation}
where $\bar{\mathcal{S}}_{\aopd\aop}(\omega)$ and $\bar{\mathcal{S}}_{\aop\aopd}(\omega)$ are the one-sided emission and absorption mechanical spectra, in the actual protocol.
\begin{equation}
    \begin{aligned}
        \bar{\mathcal{S}}_{\op{L}_i^\dag \op{L}_i}(\omega) & =  \frac{(\tilde{n}_{\text{th}} + \delta_{ei}) \tilde{\kappa}_{m}}{(\omega - \Delta)^2 + (\tilde{\kappa}_{m}/2)^2},
    \end{aligned}
\end{equation}
where $i\in\{g,e\}$ and $(\tilde{n}_{\text{th}}, \tilde{\kappa}_{m}) = (n_{\text{th},p}',\kappa_{m,p}')$ with $p\in \{g,e, ge\}$ based on the used protocol.

\subsection{Effect of the qubit imperfections}
\label{app:sideband_asymmetry_imperfect}
In this section, we account for imperfect state preparation and readout, as well as intrinsic qubit decoherence.
We assume the qubit is initialized in \(\ket{g}\) and \(\ket{e}\) with preparation fidelities \(\eta_g\) and \(\eta_e\), respectively.
The qubit’s energy relaxation/excitation (at rate \(\kappa_1/2\)) and pure dephasing (at rate \(\kappa_\phi\)) are described by the Lindblad jump operators
\begin{equation}
    \op{L_{q,\downarrow}} = \sqrt{\frac{\kappa_1}{2}}\,\op{\sigma},\quad
    \op{L_{q,\uparrow}}   = \sqrt{\frac{\kappa_1}{2}}\,\op{\sigma}^\dag,\quad
    \op{L_{q,\phi}}      = \sqrt{\frac{\kappa_\phi}{2}}\,\op{\sigma_z}.
\end{equation}
The relevant hierarchy in our experimental regime is \(\Delta, \kappa_m n_{th} \ll \Omega\sqrt{n_{th}} \ll \kappa_1, \kappa_\phi \).
Therefore, we can perform a perturbation treatment on the losses of the qubit.
The Lindblad super-operator, \(\sop{L}\) and the joint density matrix, \(\rhop\), can be vectorized using the column-stacking convention on the qubit space.
The total Hilbert space is \(\mathcal{H}_{q}\otimes\mathcal{H}_m\), then the vectorized density matrix reads \(\vec{\rhop} = (\op{\rho_{ee}}, \op{\rho_{eg}}, \op{\rho_{ge}}, \op{\rho_{gg}})^T\), where each elements is an operator on \(\mathcal{H}_m\).
An operator \(\op{O}\) acting on the left (resp.\ right) of the oscillator density matrix is written \(\opr{O} \) (resp.\(\opl{O} \)).
Using these conventions, the super-operators can be written as, \(\sop{L} = \sop{L}_0 + \sop{L}_1 + \sop{L}_m \), with
\begin{equation*}
    \sop{L}_0 =
    \begin{pmatrix}
        -\kappa_1/2 & 0         & 0         & \kappa_1/2  \\
        0           & -\kappa_2 & 0         & 0           \\
        0           & 0         & -\kappa_2 & 0           \\
        \kappa_1/2  & 0         & 0         & -\kappa_1/2
    \end{pmatrix},
\end{equation*}
capturing the qubit loss operators and
\begin{equation*}
    \sop{L}_1 = \frac{\Omega}{2}
    \begin{pmatrix}
        0             & \opr{a}^\dag  & \opl{a}  & 0            \\
        -\opr{a}      & 0             & 0        & \opl{a}      \\
        -\opl{a}^\dag & 0             & 0        & \opr{a}^\dag \\
        0             & -\opl{a}^\dag & -\opr{a} & 0
    \end{pmatrix}
\end{equation*}
capturing the qubit membrane coupling.

\tocless{\subsubsection}{Perturbative treatment}
As \(\sop{L}_0\) acts only on the qubit and \(\sop{L}_m\) only on the mechanical mode, \([\sop{L}_0, \sop{L}_m] = 0\).
Under the approximation of a weak coupling, the evolution super-operator can be approximated by the truncated Dyson series
\begin{widetext}
    \begin{equation}
        \begin{aligned}
            e^{\sop{L}t}  \simeq
                   & e^{(\sop{L}_0 + \sop{L}_m)t} + \int_0^t \!\! dt_1 \;e^{(\sop{L}_0 + \sop{L}_m)(t-t_1)}\sop{L}_1 e^{(\sop{L}_0 + \sop{L}_m) t_1}                    + \int_0^t \!\! dt_2 \!\! \int_0^{t_2} \!\! dt_1 \;e^{(\sop{L}_0 + \sop{L}_m)(t-t_2)}\sop{L}_1 e^{(\sop{L}_0 + \sop{L}_m) (t_2-t_1)}\sop{L}_1 e^{(\sop{L}_0 + \sop{L}_m) t_1} \\
            \simeq & e^{\sop{L}_m t} \left( e^{\sop{L}_0t} + \int_0^t dt_1\; e^{\sop{L}_0(t-t_1)}\sop{L}_1 e^{\sop{L}_0t_1} + \int_0^t dt_2\int_0^{t_2} dt_1 \;e^{\sop{L}_0(t-t_2)}\sop{L}_1 e^{\sop{L}_0 (t_2-t_1)}\sop{L}_1 e^{\sop{L}_0 t_1}\right),
        \end{aligned}
        \label{eq:dyson_series_definition}
    \end{equation}
\end{widetext}
where we used both that \(\sop{L}_0\) and \(\sop{L}_m\) commutes and the Baker-Campbell-Hausdorff formula to factorize \(e^{\sop{L}_m t}\).
Upon exponentiation of \(\sop{L}_0\), we get
\begin{equation}
    e^{\sop{L}_0t} =
    \begin{pmatrix}
        \frac{1+e^{-\kappa_1t}}{2} & 0               & 0               & \frac{1-e^{-\kappa_1t}}{2} \\
        0                          & e^{-\kappa_2 t} & 0               & 0                          \\
        0                          & 0               & e^{-\kappa_2 t} & 0                          \\
        \frac{1-e^{-\kappa_1t}}{2} & 0               & 0               & \frac{1+e^{-\kappa_1t}}{2}
    \end{pmatrix}.
\end{equation}
The first integral in \refeq{eq:dyson_series_definition}, can then be computed as
\begin{equation}
    \sop{I}_1(t) \equiv \frac{\Omega}{2}
    \begin{pmatrix}
        0            & \aop_+^\dag  & \aop_-  & 0           \\
        -\aop_+      & 0            & 0       & \aop_-      \\
        -\aop_-^\dag & 0            & 0       & \aop_+^\dag \\
        0            & -\aop_-^\dag & -\aop_+ & 0
    \end{pmatrix},
\end{equation}
where we defined \(\aop_{\pm} = \tau_\Sigma \aop_\Sigma \pm \tau_2 \aop_\Delta \), with
\begin{gather*}
    \tau_\Sigma(t) = \frac{\kappa_1\tau_1(t) - \kappa_2\tau_2(t)}{\kappa_1 - \kappa_2},\\
    \tau_1(t) =\frac{1 - e^{-\kappa_1 t}}{\kappa_1}, \quad
    \tau_2(t) = \frac{1 - e^{-\kappa_2 t}}{\kappa_2},\\
    \aop_\Sigma = \frac{\opr{a} + \opl{a}}{2}, \quad
    \aop_\Delta = \frac{\opr{a} - \opl{a}}{2}.
\end{gather*}
Notably, \(\lim_{\kappa_2 \to \kappa_1}\tau_\Sigma(t) =  t e^{-\kappa_1 t}\) and \( \lim_{\kappa_1 \to 0}\tau_1(t)= \lim_{\kappa_2 \to 0}\tau_2(t) = t\).
Due to the matrix structure of this term, it captures the entanglement of the membrane and qubits states, but it doesn't capture the population transfer responsible for heating or cooling of the membrane.
In order to capture such effect, we have to consider the second order perturbation term of \refeq{eq:dyson_series_definition}, that can be written as
\begin{equation}
    \sop{I}_2(t) = \frac{\Omega^2}{4}
    \begin{pmatrix}
        \op{V}_{1, 1} & 0             & 0             & \op{V}_{1, 4} \\
        0             & \op{V}_{2, 2} & \op{V}_{2, 3} & 0             \\
        0             & \op{V}_{3, 2} & \op{V}_{3, 3} & 0             \\
        \op{V}_{4, 1} & 0             & 0             & \op{V}_{4, 4}
    \end{pmatrix}
\end{equation}
where \(\{\op{V}_{i, j}\}_{i, j\in [\![1, 4]\!]}\) terms are not explicitly written for simplicity.

\tocless{\subsubsection}{Kraus operators}

The Kraus operators acting on the membrane degree of freedom can be computed as
\begin{equation}
    \begin{aligned}
         & \sop{K}_{\pm, i}\rhop_m = \mathrm{Tr}_q[\ket{\pm}\bra{\pm}e^{\sop{L}\tint}(\rhop_m\otimes\rhop_{q, i})] \\
         & \text{with }
        \rhop_{q, i} =
        \begin{pmatrix}
            1 - p_i & 0 & 0 & p_i
        \end{pmatrix}^T
    \end{aligned}
\end{equation}
where \(p_g = (1+\eta_g)/2\) and \(p_e = (1-\eta_e)/2\).
They can be approximated using \refeq{eq:dyson_series_definition} as
\begin{equation}
    \begin{aligned}
        \sop{K}_{\pm, i} = & \frac{e^{\sop{L}_m\tint}}{2}\Bigl( \II \pm \frac{\Omega}{4}((2p_i-1)\tau_\Sigma + \tau_2) \sop{M}[\aop]                                  \\
                           & \pm \frac{\Omega}{4}((2p_i-1)\tau_\Sigma - \tau_2)\sop{M}[\aopd]                                                                         \\
                           & + \frac{\Omega^2}{4}\left(\frac{\tint - \tau_2}{\kappa_2} + (2p_i-1)\frac{\tau_2 - \tau_1}{\kappa_1 - \kappa_2}\right)\sop{D}[\aop]      \\
                           & + \frac{\Omega^2}{4}\left(\frac{\tint - \tau_2}{\kappa_2}-(2p_i-1)\frac{\tau_2 - \tau_1}{\kappa_1 - \kappa_2}\right)\sop{D}[\aopd]\Bigr)
    \end{aligned}
\end{equation}
Due to the limited qubit lifetime, the action of a qubit prepared in \(\ket{g}\) (resp. \(\ket{e}\) also weakly measures the \(\aopd\) (resp. \(\aop\)) operator.

To account for the readout infidelity, we'll write the readout confusion matrix in terms of the single-shot assignment error probabilities \(\varepsilon_g = 1-F_g\) and \(\varepsilon_e = 1-F_e\).
As the state is readout in the \(\ket{\pm}\) basis with a \(\sqrt{\op{X}}\) pulse, the readout errors on state \(\ket{+}\) (resp. \(\ket{-}\)) are the one of \(\ket{e}\) (resp. \(\ket{g}\)), defining \(\varepsilon_{+/-} = \varepsilon_{e/g}\).
The effective Kraus operator acting on the membrane can be written as
\begin{equation}
    \bar{\sop{K}}_{\pm, i} = (1-\varepsilon_\pm) \sop{K}_{\pm, i} + \varepsilon_\pm \sop{K}_{\mp, i}.
\end{equation}

\tocless{\subsubsection}{Emission and absorption spectra}
We can compute all the previous steps to compute the autocorrelation we applied to the lossless qubit, except for the case where \(k=0\).
Indeed, as the readout of the qubit is asymmetric, the autocorrelation at \(k=0\) is modified to \(\mathcal{C}_k^i = 1 - (\varepsilon_g - \varepsilon_e)^2\).
Computing the spectra from this autocorrelations, we obtain that
\begin{equation}
    \begin{aligned}
        \mathcal{S}_{mm}^g(\omega_k') = & (1-\varepsilon_g - \varepsilon_e)^2\eta_g\frac{\Omega^2\tau_\Sigma^2}{4}
        \Bigl(\frac{\eta_g + \tau_2/\tau_\Sigma}{2}\bar{\mathcal{S}}_{\aopd\aop}(\omega_k') +                                                                                                             \\
                                        & \frac{\eta_g - \tau_2/\tau_\Sigma}{2}\bar{\mathcal{S}}_{\aop\aopd}(\omega_k'))
        \Bigr) + \tcycle(1- (\varepsilon_g - \varepsilon_e)^2),                                                                                                                                           \\
        \mathcal{S}_{mm}^e(\omega_k') = & (1-\varepsilon_g - \varepsilon_e)^2\eta_e\frac{\Omega^2\tau_\Sigma^2}{4} \Bigl(\frac{\eta_e - \tau_2/\tau_\Sigma}{2} \bar{\mathcal{S}}_{\aopd\aop}(\omega_k') + \\
                                        & \frac{\eta_e + \tau_2/\tau_\Sigma}{2}\bar{\mathcal{S}}_{\aop\aopd}(\omega_k'))
        \Bigr) + \tcycle(1- (\varepsilon_g - \varepsilon_e)^2).                                                                                                                                           \\
    \end{aligned}
    \label{eq:s_mm_final_spectrum}
\end{equation}
Using the amplitude of the measured thermal peak for the preparation in \(\ket{g}\) at 10~mK in \reffig{fig:spec_asym}, we estimate \(\Omega/2\pi = 1.50(8)\)~kHz.

The normalized spectra are expressed in units of a calibration peak amplitude resulting from a weak drive on the qubit, the resulting sinus cardinal amplitudes in the spectra are proportionate to \((1-\varepsilon_g-\varepsilon_e)^2\eta_i^2\).
The area of the normalized Lorentzian peaks therefore reads
\begin{equation}
    \begin{aligned}
        A_g = & \frac{\Omega^2\tau_\Sigma^2}{4}\left(\nmth + \frac{1}{2}-\frac{\tau_2}{2\eta_g\tau_\Sigma}\right) \\
        A_e = & \frac{\Omega^2\tau_\Sigma^2}{4}\left(\nmth + \frac{1}{2}+\frac{\tau_2}{2\eta_e\tau_\Sigma}\right)
    \end{aligned}
\end{equation}
Performing a linear fit of the areas as a function of the cryostat temperature, we are able to express the amplitude in term of phonon number,
\begin{equation}
    \begin{aligned}
        \bar{n}_g = & \nmth + \frac{1}{2}-\frac{\tau_2}{2\eta_g\tau_\Sigma} \\
        \bar{n}_e = & \nmth + \frac{1}{2}+\frac{\tau_2}{2\eta_g\tau_\Sigma}
    \end{aligned}
\end{equation}
The expected asymmetry of the spectrum used in \reffig{fig:spec_asym} writes
\begin{equation}
    \bar{n}_e - \bar{n}_g = \frac{\tau_2}{\tau_\Sigma}\frac{\eta_g+\eta_e}{2\eta_g\eta_e}.
\end{equation}

\tocless{\subsection}{Data analysis of the spectrum analyzer}
\label{sec:data_analysis}

In order to estimate the noise spectral density $S_{mm}^{g/e}$ of the qubit telegraphic signal, we first group the measurement outcomes $\{m_k\}$ into batches of $N = 10\,000$ samples (see \reffig[(d)]{fig:spectrum_analyzer}).
For each batch, we compute the discrete Fourier transform, $\mathcal{F}[m_k]_r$, where the frequency bins are defined as
\begin{equation}
    \omega_r = 2 \pi r/N\tcycle \quad\text{for}\quad 0\leq r<N.
\end{equation}
We then average the squared magnitudes over all batches to obtain the noise spectral density
\begin{equation}
    \label{eq:averaging}
    S_{mm}(\omega_r) = \overline{|\mathcal{F}[m_k]_r|^2}\tcycle/N,
\end{equation}
with the overline indicating an average over batches corresponding to distinct classical trajectories \(\{x(t_k)\}_{0\leq k<N}\) (see \reffig[(e)]{fig:spectrum_analyzer}).
At our sampling rate, the aliasing-free frequency span is $\omega_{N/2} \approx  2\pi \times 32~$kHz.
We therefore choose a small qubit-membrane detuning $\Delta/2\pi = 5~$kHz to ensure the oscillations of $x(t)$ remain entirely within the Nyquist zone.

The spectra displayed in \reffig[(e–f)]{fig:spectrum_analyzer} and \reffig[(c)]{fig:spec_asym} are obtained from $N_{\rm batches} \sim \mathrm{1000}$ and $\mathrm{4\times10^5}$ Fourier‐transformed batches.
To determine the statistical error on the estimated spectrum, we use a bootstrap approach: the averaging described by \refeq{eq:averaging} is repeated over 1000 different datasets.
Each dataset is obtained from the original data by drawing a resample with replacement of the $N_{\rm batches}$ Fourier‐transformed batches.
The 1000 different spectra are then accumulated in histograms, represented by the shaded regions in \reffig[(e–f)]{fig:spectrum_analyzer} and \reffig[(c)]{fig:spec_asym}.
The quoted errors on the fitted linewidths $\kappa_g'$, $\kappa_e'$ and peak areas $A_g$, $A_e$, as well as the error bars in~\reffig[(d–e)]{fig:spec_asym}, are the standard deviations of those 1000 fits.

Since the mechanical peak area scales as $V_\text{offset}^2$, any slow drift in $V_\text{offset}$ during a monotonic temperature sweep (see Sec. (S2.5) of supplementary information) would introduce a spurious temperature dependence in the sideband-asymmetry data of \reffig[(c)-(d)]{fig:spec_asym}.
To avoid this, we interleaved the measurements by cycling the cryostat through each setpoint (2 h per point) for 16 complete cycles, so that any residual offset drift acts uniformly across all temperatures. \begin{figure*}[p]
    \centering
    \includegraphics[width=0.6\textwidth]{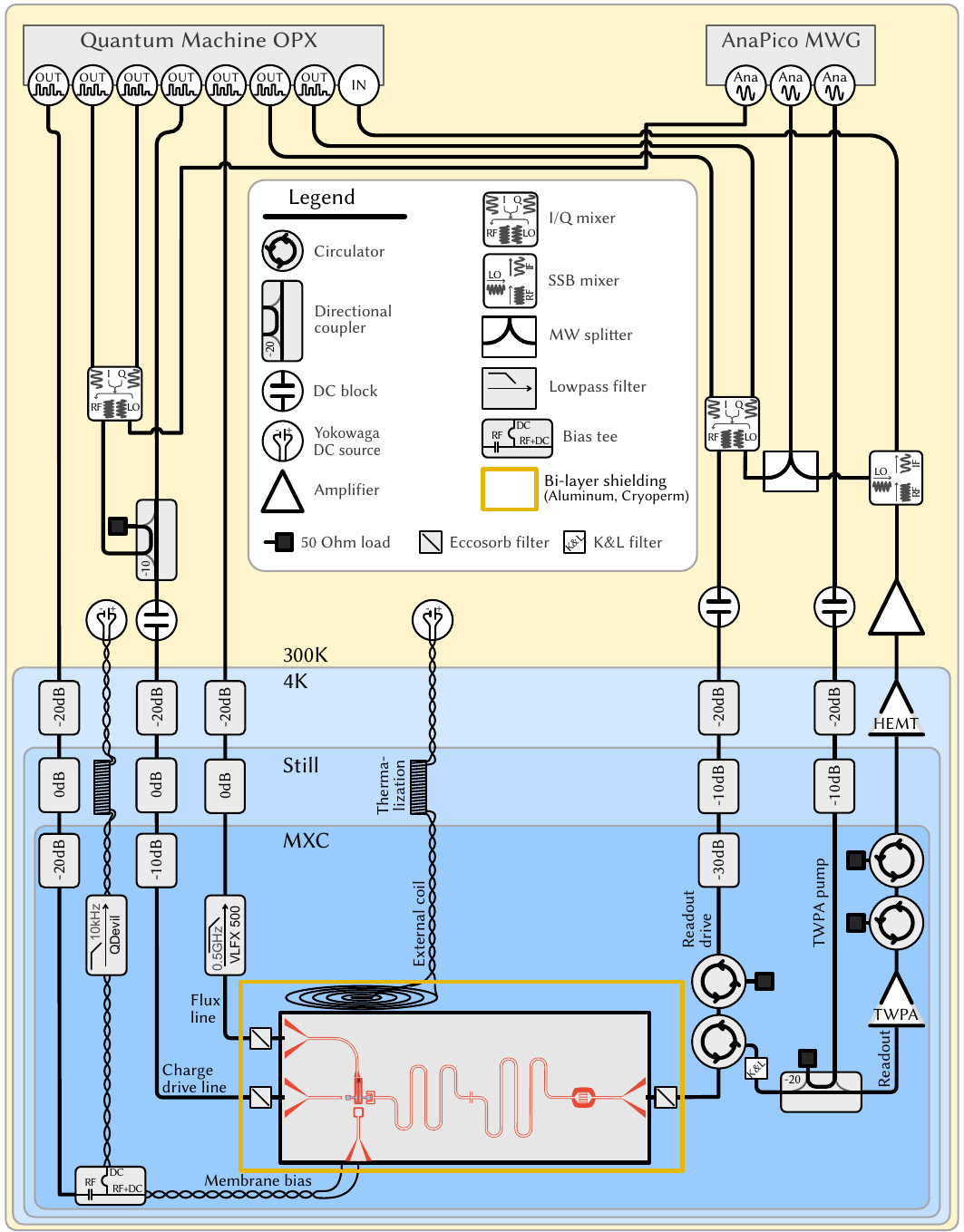}
    \caption{
        A detailed wiring schematic of the experimental setup, showing both the RF and DC lines inside the cryostat and the control electronics at room temperature. `Yoko'
        refers to the \textit{Yokogawa} DC source, `QM` to the \textit{Quantum Machine} FPGA, and `Ana` to the \textit{Anapico} microwave source.}
    \label{fig:cryostat_setup}
\end{figure*}

\section{Experimental Setup}
\label{sec:experimental_setup}

The experimental setup is shown in \reffig{fig:cryostat_setup}.
RF signals are controlled and measured using a heterodyne detection scheme, implemented via an FPGA-based controller (\textit{OPX, Quantum Machine}) combined with a microwave source (\textit{Anapico APUASYN20-4}). The magnetic flux through the qubit is tuned using a \textit{Yokogawa 7651} DC source, which supplies current to a superconducting coil placed directly to the chip enclosure. Fast flux tuning is achieved via an on-chip flux line.
Signal amplification is performed using a traveling-wave parametric amplifier (TWPA), a high-electron-mobility transistor (HEMT) amplifier, and a room-temperature amplifier. The DC membrane bias is controlled by a \textit{Yokogawa GS200} source and filtered through a commercial 10~kHz low-pass filter (\textit{QDevil}) to suppress high-frequency noise.

\clearpage

 \bibliographystyleS{naturemag}
{\let\oldaddcontentsline\addcontentsline
    \renewcommand{\addcontentsline}[3]{}\bibliographyS{bib-supp}
    \let\addcontentsline\oldaddcontentsline
}

\end{document}